\documentclass[useAMS,usenatbib]{mnras}
\usepackage{aas_macros}
\usepackage{hyperref}
\usepackage{cleveref}
\usepackage{graphicx}
\usepackage{amsmath,amssymb}
\usepackage[T1]{fontenc}
\usepackage{aecompl}
\title[Jet Heating in Galaxy Clusters]{Simulations of Jet Heating in Galaxy Clusters: Successes and  Challenges}
\author[D. Martizzi et al.]{\parbox[t]{\textwidth}{Davide Martizzi$^{1,2}$\thanks{E-mail: davide.martizzi@nbi.ku.dk}, Eliot Quataert$^{3}$, Claude-Andr\'{e} Faucher-Gigu\`{e}re$^{4}$, Drummond Fielding$^{3}$}\vspace*{6pt}\\
$^{1}$Dark Cosmology Centre, Niels Bohr Institute, University of Copenhagen, 2100 Copenhagen, Denmark. \\
$^{2}$Department of Astronomy and Astrophysics, University of California, Santa Cruz, CA 95064, USA. \\
$^{3}$Department of Astronomy and Theoretical Astrophysics Center, University of California, Berkeley, CA 94720-3411, USA. \\
$^{4}$Department of Physics \& Astronomy and Center for Interdisciplinary Exploration and Research in Astrophysics (CIERA),\\ Northwestern University, Northwestern University, Evanston, IL 60208-3112, USA.\\
}

\begin{document}

\maketitle

\label{firstpage}

\begin{abstract}
We study how jets driven by active galactic nuclei influence the cooling flow in Perseus-like galaxy cluster cores with idealised, non-relativistic, hydrodynamical simulations performed with the Eulerian code {\sc athena} using high-resolution Godunov methods with low numerical diffusion. We use novel analysis methods to measure the cooling rate, {the heating rate associated to multiple mechanisms}, and {the power associated with adiabatic compression/expansion. A significant reduction of the cooling rate and cooling flow within 20 kpc from the centre can be achieved with kinetic jets. However, at larger scales and away from the jet axis, the system relaxes to a cooling flow configuration. Jet feedback is anisotropic and is mostly distributed along the jet axis, where the cooling rate is reduced and a significant fraction of the jet power is converted into kinetic power of heated outflowing gas}. Away from the jet axis {weak shock heating represents the dominant heating source}. Turbulent heating is significant only near the cluster centre, but it becomes inefficient at $\sim 50$~kpc scales where it only represents a few percent of the total heating rate. Several details of the simulations depend on the choice made for the hydro solver, a consequence of the difficulty of achieving proper numerical convergence for this problem: current physics implementations and resolutions do not properly capture multi-phase gas that develops as a consequence of thermal instability. These processes happen at the grid scale and leave numerical solutions sensitive to the properties of the chosen hydro solver.
\end{abstract}

\begin{keywords}
galaxies: clusters: general -- galaxies: active -- galaxies: jets -- methods: numerical
\end{keywords}

\section{Introduction}

Galaxy clusters reside in the most massive dark matter halos in the Universe and contain large reservoirs of hot plasma, the intracluster medium (ICM). A significant fraction of galaxy clusters exhibit the so-called cool-core configuration in which gas at the cluster centre has cooling times $\lesssim 1$~Gyr and low entropy. In these conditions strong cooling flows should develop \citep{1994ARA&A..32..277F}, leading to the accumulation of large quantities of cold gas in cluster cores and to the formation of massive gas-rich central galaxies \citep{2012MNRAS.420.2859M, 2013MNRAS.436.1750R, 2014MNRAS.443.1500M}. However, such extreme cooling flows are not observed \citep{1997MNRAS.292..419W, 1998Ap&SS.263...83C, 2000MNRAS.315..269A, 2001MNRAS.326.1228B, 2003ApJ...590..207P, 2008MNRAS.385.1186S}.

The discrepancy between observed and theoretically predicted cooling flows in cool-core clusters led theorists to postulate the existence of heating mechanisms in cluster cores that regulate cooling flows. Although thermal conduction can influence the 
thermodynamics of cluster cores \citep{1989ApJ...338..761R, 1998PhRvL..80.3077C, 2001ApJ...562L.129N, 2002ApJ...581..223R, 2004MNRAS.347.1130V}, it has been shown not to be sufficient to explain the discrepancy on its own 
\citep{2000MNRAS.317L..57E, 2003ApJ...582..162Z, 2004MNRAS.347.1130V, 2004ApJ...606L..97D, 2009ApJ...703...96P}. Theoretical studies of the co-evolution of galaxies and supermassive black holes (SMBHs) have suggested that active galactic nuclei (AGN) may provide 
a significant source of heating that may regulate the growth of central galaxies in massive dark matter halos \citep{1993MNRAS.263..323T, 1997ApJ...487L.105C, 1998A&A...331L...1S}. Extensive theoretical and numerical work has been performed to demonstrate the 
feasibility of this solution to the cooling flow problem \citep{2002MNRAS.332..729C, 2002Natur.418..301B, 2006MNRAS.365...11C, 2007MNRAS.380..877S, 2009MNRAS.398...53B, 2012MNRAS.420.2662D}. On the observational side, significant evidence has been also produced 
that AGN provide heating in cluster cores via their jets that pierce through the ICM, inject highly relativistic particles into the ICM, inflate large, hot bubbles, and excite shocks and turbulent motions in cluster cores \citep{2007ARA&A..45..117M, 2012ARA&A..50..455F, 2014Natur.515...85Z}. 

Cosmological zoom-in hydrodynamical simulations of galaxy clusters including AGN feedback are generally more successful at reproducing cluster properties than simulations that do not include this effect \citep{2007MNRAS.380..877S, 2010MNRAS.409..985D, 2011MNRAS.414..195T, 2014MNRAS.441.1270L, 2014MNRAS.443.1500M, 2015ApJ...813L..17R, 2017MNRAS.470..166H, 2017MNRAS.470.4186B}. AGN feedback is also implemented in all recent major large volume cosmological hydrodynamical simulations \citep{2014MNRAS.444.1518V, 2015MNRAS.446..521S, 2016MNRAS.463.3948D, 2016MNRAS.462.3265D, 2017MNRAS.465.2936M, 2018MNRAS.473.4077P}. Unfortunately, cosmological simulations are useful for predicting the global properties of galaxy clusters and their galaxies, but fall short in terms of resolution. Furthermore, the evolution of cluster cores in cosmological simulations is very non-linear, which complicates the analysis of heating from AGNs. {For this reason, several authors have preferred to complement the knowledge gained from cosmological simulations with results from idealised simulations of AGN jet heating: some of the previous work focused on non-precessing jets 
\citep{2004MNRAS.348.1105O, 2007MNRAS.376.1547C, 2011MNRAS.411..349G, 2017MNRAS.469.4148C, 2017MNRAS.470.4530W, 2018MNRAS.473.1332G}, whereas other work studied precessing jets \citep{2012ApJ...746...94G, 2012ApJ...747...26L, 2014ApJ...789...54L, 2016ApJ...829...90Y, 2017MNRAS.472.4707B}, jets with large opening angle \citep{2015ApJ...811..108P, 2017ApJ...845...91H, 2018arXiv180100408H}, or spherical AGN `energy dumps' \citep{2015ApJ...815...41R} which promote the redistribution of the heating rate over large volumes. }

Recent theoretical models and idealised simulations established the importance of thermal instabilities developing in the ICM when jet heating is introduced \citep{2012MNRAS.420.3174S, 2012MNRAS.419.3319M, 2012ApJ...747...26L, 2014ApJ...789...54L, 2015ApJ...808...43M, 2015Natur.519..203V}. Such instabilities lead to the formation of clumps and filaments of cold gas embedded in the hotter ICM. The existence of such multi-phase sub-structure is expected to influence the cycle of activation and deactivation of a jet and the observational properties of the baryons in cluster cores \citep{2015Natur.519..203V}. It is very important for simulations of jet heating in galaxy clusters to capture the onset of this instability, which can be achieved only at sufficiently high spatial resolution ($\lesssim 1$ kpc). 

The combination of Eulerian numerical methods with the high numerical resolution achievable on modern supercomputers allowed several groups to perform detailed analysis of the balance between heating and cooling in cool-core clusters \citep{2014ApJ...789...54L, 2016ApJ...829...90Y, 2017ApJ...847..106L, 2017ApJ...841..133M}. In this paper, we extend this work and offer a complementary analysis with independent techniques. We perform idealised simulations of jet heating with the {\sc athena} code \citep{2008ApJS..178..137S}. The spatial resolution of our simulations is $\sim 200$~pc, comparable to or better than that reached in previous work \citep[][]{2017ApJ...841..133M, 2016ApJ...829...90Y}, but our setup differs in the choice of numerical and analysis techniques. On the numerical side, our fiducial simulations adopt a numerical solver with much lower numerical diffusion than in most of the recent literature. On the analysis side, we carefully estimate the cooling and heating rates of the gas cell-by-cell, which allows a very detailed characterization of the state of the system throughout its evolution in time. In particular, the total heating rate is measured on-the-fly using a Lagrangian entropy tracer \citep{2015MNRAS.454.1848R}. Equipped with these tools, we want to (I) characterise how the cooling flow is regulated by AGN feedback, (II) identify the main heating mechanism in different regions of the cluster core, (III) assess the robustness of the results and compare to previous work performed with different hydrodynamical methods. 

The paper is structured as following: Section 2 discusses the details of our numerical setup and analysis methods; Section 3 shows our main results and discusses them in the context of previous work; Section 4 summarises the paper and our conclusions. 

\section{Simulation Setup}
\label{sec:setup}

We carry out 3-d hydrodynamical simulations with the {\sc athena} code \citep{2008ApJS..178..137S}, with the goal of studying AGN jet heating in the core of 
massive galaxy clusters that develop a strong cooling flow. {\sc athena} offers a variety of solvers and integrators for the equations of ideal, compressible, hydrodynamics on Cartesian grids. Additionally, the code offers the possibility of statically refining the computational domain (Static Mesh Refinement, SMR) to achieve higher resolution. SMR allows the user to explicitly define the refinement strategy in advance and to achieve great control on the number of resolution elements and on load balancing when the code is run on a large number of cores. 

Several Riemann solvers are implemented in {\sc athena}, including the Roe solver, the Harten-Lax-van Leer-Contact (HLLC) solver and the Harten-Lax-van Leer-Einfeldt (HLLE) solver. We encountered significant difficulties when setting up simulations with the Roe solver, which is the most accurate and the one with the lowest numerical diffusion. Simulations with the Roe solver introduce significant numerical errors when clouds with high density contrast crossed coarse-fine boundaries in the SMR grid, which lead to spurious heating in the solution in most cases. HLLC and HLLE both offer fast, approximate solutions to the Riemann problem. For most of our simulations we adopt the HLLC solver, but we also experiment with HLLE to check the robustness of our results. HLLC efficiently captures shocks and contact discontinuities, but is somewhat more diffusive than the Roe solver. HLLE has higher numerical diffusion and it is known not to capture contact discontinuities, which often develop at the interface of hot and cold structures in multi-phase media. After performing tests with Roe, HLLE and HLLC, we conclude that the latter offers more stable solutions and is affected by fewer numerical artifacts for the problem we solve in this paper, as we show and discuss in Subsections \ref{sec:kin_therm} and \ref{sec:rsolvers}. 

{\sc athena} also offers different choices for the time integration of the fluid equations. The Corner Transport Upwind (CTU) integrator is chosen for our simulations with the HLLC Riemann solver. CTU offers higher accuracy and smaller numerical diffusion compared to the Van Leer (VL) integrator. We adopt VL to integrate the fluid equations performed with the HLLE Riemann solver. Therefore, our simulations come in two flavors: HLLC+CTU (more accurate, lower numerical diffusion) and HLLE+VL (less accurate, higher numerical diffusion). 

Finally, we perform second-order (piecewise linear) reconstruction of the hydrodynamic variables. It is important to stress that our choices for Riemann solver and integrator have inherently {\it lower numerical diffusion} compared to other schemes previously adopted in the literature for similar simulations \citep[e.g. {\sc zeus}, ][]{2012ApJ...747...26L, 2017ApJ...847..106L, 2017ApJ...841..133M}. In the rest of the paper, we will discuss the importance of this difference.

{We consider runs with several different setups, which label using the format ${\rm XXX\_YYY\_ZZZ}$, where XXX is a label for the resolution, YYY is a label for the jet physics, and ZZZ is a label for the combination of Riemann solver and time integrator. We perform simulations at three resolutions XXX = LR (low resolution), MR (medium resolution), HR (high resolution), respectively. We use three setups for the jet physics YYY = COOL (cooling only, no jet), JET (pure kinetic energy injection) and MIXED (mixed thermal and kinetic energy injection), respectively. Finally, we use two choices for the Riemann solver/time integrator label, ZZZ = HLLC (for HLLC solver + CTU integrator) and ZZZ = HLLE (HLLE solver + VL integrator). The runs considered in this paper are summarised in Table~\ref{tab:list_sims}.}

\subsection{Refinement Scheme}

Since we focus on the heating by AGN jets in cluster cores, we only simulate the central region of a massive ($M\sim 10^{15}$ M$_{\odot}$) Perseus-like cluster. For this reason, all our 
simulations adopt a cubic box of side $L=400$ kpc. The centre of the cluster is placed at the box centre and SMR is adopted to achieve high resolution in the regions influenced by jet heating. 
In all runs SMR is implemented by nesting multiple concentric cubic meshes. 
Each refined zone at a given level is a cube of side half of that of the coarser level. The number of resolution elements in each level of refinement is kept constant, 
so that the spatial resolution doubles by passing from one zone to one with higher level of refinement. The LR (low resolution) runs adopt a root grid plus 1 level of refinement, with $128^3$ elements per grid, 
reaching an effective resolution $\Delta x$ = 1.562 kpc at the highest level of refinement. The MR (medium resolution) runs adopt a root grid plus 3 levels of SMR, with $128^3$ elements per grid, 
reaching an effective resolution $\Delta x$ = 390 pc at the highest level of refinement. The HR (high resolution) runs also use a root grid plus 3 levels of SMR, but each grid has a $256^3$, i.e. the resolution is doubled 
everywhere with respect to MR, and the effective resolution is $\Delta r$ = 195 pc.

\subsection{Initial and Boundary Conditions}

Our numerical simulations are based on spherically symmetric, semi-analytical initial conditions. The model assumes the ICM to be in hydrostatic equilibrium under the influence of a static, external 
Navarro-Frenk-White (NFW) gravitational potential associated with a dark matter halo. The gravitational potential is given by the standard NFW formula \citep{1997ApJ...490..493N}:
\begin{equation}\label{eq:pot}
\Phi(x)=4\pi G r_{\rm s}^2 \rho_{\rm s} \frac{\ln (1+\frac{r}{r_{\rm s}})}{\frac{r}{r_{\rm s}}}
\end{equation}
where $\rho_{\rm s}$ is a characteristic density of the halo, $r_{\rm s}$ is the halo scale radius. 
Let $r_{\rm 200}$ be the radius within which the mean density is 200 times the critical density, then we can define 
the concentration
\begin{equation}
 c_{\rm 200}=\frac{r_{\rm 200}}{r_{\rm s}}.
\end{equation}
We also label the mass within $r_{\rm 200}$ as $M_{\rm 200}$. 
Once concentration $c_{\rm 200}$ and halo mass $M_{\rm 200}$ are set, the NFW potential is uniquely determined. 
However, observational measurements and cosmological N-body simulations have shown that a well established 
mass-concentration exists for halos in a broad range of halo masses 
\citep{2014MNRAS.441.3359D, 2014MNRAS.441..378L, 2015ApJ...806....4M, 2016MNRAS.457.4340K}. 
We adopt the mass-concentration relation at redshift $z=0$ from \cite{2014MNRAS.441.3359D}:
\begin{equation}
 c_{\rm 200} = 8.03\times\left(\frac{M_{\rm 200}}{10^{12} \hbox{ M}_{\odot}/h}\right)^{-0.101}.
\end{equation}
We set $h = 0.7$ for the Hubble parameter. Once the mass-concentration relation is set, the NFW 
potential model only depends on the choice of $M_{\rm 200}$. 

Since the spatial resolution of our simulations is enough to resolve the inner $10-20$ kpc of the cluster, which in real 
clusters are dominated by the gravitational potential of the brightest cluster galaxy (BCG), we also include its 
contribution. We adopt a spherically symmetric model for the BCG mass profile that has been recently used 
by \cite{2017ApJ...841..133M}: 
\begin{equation}\label{eq:BCG}
M_{*} (<r) = M_4 \left[ \frac{(r/4\, {\rm kpc})^{0.1} (1+r/4\, {\rm kpc})^{1.33}}{2^{1.43}} \right],
\end{equation}
where $M_4$ is the BCG stellar mass within 4 kpc. 

{We do not include self-gravity of the ICM. Observational data \citep[e.g., ][]{2014MNRAS.440.2077M} 
support the fact that the ICM mass fraction at cluster-centric radii $0.05 r_{200} < r < 0.5 r_{200}$ is $\sim 0.1$. For this reason, the 
ICM gravitational potential contribution is modest. The central regions of the cluster 
at $r<50$~kpc may be different in some cases. At low redshift, the central region's potential is typically dominated by a gas poor BCG, 
which we model in equation~\ref{eq:BCG} above. However, 
for certain simulated cooling flow/merger configurations it is possible to drive large amounts of gas to the cluster centre 
whose dynamics can influence the potential at $r<10$~kpc \citep[e.g., ][]{2012MNRAS.422.3081M, 2013MNRAS.432.1947M}, 
but it is not clear how frequently these circumstances arise in real clusters. Since in our case we are interested in isolating the 
effect of the jet while keeping the rest of the physics fixed, we believe that a BCG+NFW gravitational potential is sufficient to create 
a reasonable cluster model. }

Motivated by observational evidence \citep[e.g., ][]{2006ApJ...643..730D, 2009ApJS..182...12C}, the ICM is  
assumed to have a cored entropy profile based on a Perseus-like cluster given by:
\begin{equation}
 K_{\rm gas}(x) = \frac{K_{\rm 0}}{2} \left(1+x^{\Gamma}\right),
\end{equation}
where the normalised radius is defined as  
\begin{equation}
x = \frac{r}{r_{\rm 0}},
\end{equation}
and $r_{\rm 0}$ is the characteristic size of the entropy core. $K_{\rm 0}$ is the entropy at $r_{0}$, whereas 
the central entropy is $K_{\rm 0}/2$. $\Gamma$ is the slope of the entropy profile outside the core region. 

Once entropy and external gravitational potential are determined, setting the density profile automatically determines the pressure profile. In fact, 
entropy is related to pressure and density:
\begin{equation}
 K_{\rm gas}(x) = \frac{P_{\rm }(x)}{\rho_{\rm gas}(x)^{\gamma}}
\end{equation}
where $\gamma=5/3$ is the polytropic index of the gas. Inserting this expression in the equation of hydrostatic equilibrium, a differential 
equation for the density profile is obtained:
\begin{equation}
\frac{K_{\rm 0}}{2}\frac{d}{dx} \left[ \left( 1+ x^{\Gamma} \right) \rho_{\rm gas}(x)^{\gamma} \right] = -\rho_{\rm gas}(x)\frac{d\Phi}{dx}.
\end{equation}
We numerically solve the latter with a 4th-order Runge-Kutta scheme with boundary condition 
$\rho_{\rm gas}(x=1)=\rho_0$. Once the density profile $\rho_{\rm gas}(x)$ is known, the gas pressure is computed:
\begin{equation}
 P_{\rm }(x)=K_{\rm gas}(x)\rho_{\rm gas}(x)^{\gamma}
\end{equation}
The temperature profile is:
\begin{equation}
 T_{\rm gas}(x)=\frac{\mu m_{\rm p}}{k_{\rm B}}\frac{P_{\rm gas}(x)}{\rho_{\rm gas}(x)}.
\end{equation}

We do not include rotation of the ICM component in the presented simulations, but we have experimented with it. Our tests suggest that the results including ICM 
rotation are qualitatively similar to those reported below, therefore we omit these test from this paper. 

With our choices, the model for the initial condition is uniquely determined by parameters $M_{\rm 200}$, $M_4$, $r_{0}$, $K_{\rm 0}$, $\Gamma$, $\rho_{\rm 0}$. 
The goal of this study is to investigate jet heating in massive, cool-core clusters. For this reason, we set the parameters of the initial conditions to achieve 
conditions similar to those of the Perseus cluster \citep{2014MNRAS.437.3939U}. The halo mass is set to $M_{\rm 200} = 1.0\times10^{15}$ M$_{\rm \odot}$; the BCG mass within 4 kpc is set to 
$M_4 = 7.5\times10^{10}$ M$_{\odot}$; the entropy core size is set to $r_{0} = 20$ kpc;
the central entropy is set to $K_{0}/2 = 10$ keV cm$^2$; the asymptotic slope of the entropy profile at large radius is set to $\Gamma = 1.75$; the gas density at radius $r_{0}$ 
is set to $\rho_{0} = 6.67\times10^{-26}$ g cm$^{-3}$ (see Table \ref{tab:ic} for a summary). 

\begin{table}
\centering
{\bfseries Initial Conditions}
\makebox[\linewidth]{
\begin{tabular}{ll}
\hline
Parameter & Value \\
\hline
 $M_{\rm 200}$ & $1.0\times10^{15}$ M$_{\rm \odot}$ \\
 $M_{\rm 4}$ & $7.5\times10^{10}$ M$_{\rm \odot}$ \\
 $r_{0}$ & 20 kpc \\
 $K_0$ & 20 keV cm$^2$ \\
 $\Gamma$ & 1.75 \\ 
 $\rho_{0}$ & $6.67\times10^{-26}$ g cm$^{-3}$ \\
\hline
\end{tabular}
}
\caption{ Parameters of the model used for the initial conditions. }\label{tab:ic}
\end{table}

Despite our slightly different parameterization, the initial conditions are intentionally chosen to be very similar to those used by other authors 
who analysed jet heating in Perseus-like clusters in recent years \citep{2014ApJ...789...54L, 2015ApJ...811...73L, 2016ApJ...829...90Y, 
2017ApJ...847..106L, 2017ApJ...841..133M}. In particular, the initial conditions and resolution are similar to those 
of \cite{2017ApJ...841..133M}, with the exclusion of the central entropy which is lower by a factor 2 in our simulations. This translates in a shorter 
cooling time and a slightly larger cooling flow. One of the advantages of our parameterization is that it allows the user to easily vary each 
parameter and simulate a variety of systems, which will be useful for future work. 

Finally, we use outflow boundary conditions which enforce zero gradients for the conservative hydrodynamic variables (density, mass flux, total energy, flux, 
passive scalars) at the boundaries of the 
computational box. It is important to keep in mind that if the simulations are evolved long enough, large amounts of mass will transfer from large 
radii towards the central regions, as a result of the cooling flow. This implies that the boundary conditions can, in principle, influence the evolution 
of the system. The choice of our box size (400 kpc) is made to ensure that the boundary conditions cannot influence the central regions for at 
least $\sim 3$ Gyr. Furthermore, to prevent the development of numerically seeded flows at the box boundary, which can develop if gravitational 
potential gradients are combined with zero gradient boundary conditions, we implement a smooth transition between the the analytical potential 
of equation~\ref{eq:pot} and a constant function (zero force) at radius $r>180$ kpc. The transition is achieved by a power law interpolation: 
\begin{equation}
\Phi_{\rm numerical} = -(0.5|\Phi|^\alpha+0.5|\Phi_{\rm const}|^\alpha)^{1/\alpha},
\end{equation}
where $\alpha \geq 5$ sets the sharpness of the transition and $\Phi_{\rm const}=\Phi({\rm r=180\, kpc})$. The net effect is the creation of a buffer region with zero gravitational force of size 43 (86) cells near the box boundaries in the 
MR (HR) case, equivalent to a physical size of $\sim 17$ kpc. 
  
\subsection{Radiative Cooling and Temperature Floor}\label{sec:cooling}

Radiative cooling is included in our simulations at each time step and uses \cite{1993ApJS...88..253S} tables for a plasma of metallicity ${\rm Z=0.3\,Z_{\rm \odot}}$. 
The cooling scheme uses sub-cycling with a 4th-order Runge-Kutta scheme to provide an accurate solution. 

Gas is not allowed to cool below a temperature floor $T_{\rm floor}$. Thermal instability arises naturally in cluster cores subject to AGN heating \citep{2012MNRAS.419.3319M, 
2012MNRAS.420.3174S, 2012ApJ...746...94G, 2014ApJ...789..153L, 2015ApJ...808...43M}, however our resolution 
is not high enough to fully resolve this process. Our numerical experiments with multiple Riemann solvers and integrators in {\sc athena} show that the numerical solution 
of the cluster heating problem is only numerically robust as long as the formation of unresolved (1-2 cells), high density clumps via thermal instability is suppressed. 
The adoption of a sufficiently high temperature floor allows us to obtain the desired effect while still allowing the cluster to develop a cooling flow (in absence of jet heating) 
and to develop resolved, high density clumps. After multiple tests, we verify that the desired behaviour is obtained for our fiducial temperature floor values, 
$T_{\rm floor}=5.0\times 10^5$ K in the MR case, and $T_{\rm floor}=2.5\times 10^5$ K in the HR case. 

Despite our temperature floor being larger than the value used by other authors in the literature \citep{2011MNRAS.411..349G, 2014ApJ...789...54L, 2015ApJ...811...73L, 
2016ApJ...829...90Y, 2017ApJ...847..106L, 2017ApJ...841..133M}, we remind the reader that the combination of HLLC solver and CTU integrator in {\sc athena} has lower 
numerical diffusion than the schemes used by the authors cited above, at the price of somewhat worse numerical stability. In tests not reported in this paper, we ran simulations with 
$T_{\rm floor}<1.0\times 10^5$ K and found that they can successfully be completed with the HLLE Riemann solver or with HLLC at LR resolution, but we encountered difficulties 
with HLLC at MR and HR resolution. Our fiducial choices for the temperature floor guarantee numerically stable solutions while keeping numerical diffusion low. 

\subsection{Gas Accretion and Jet Heating}

The key element of our simulations is the implementation of jet heating from a central supermassive black hole. Our main goal is to achieve a self-regulated modulation of the 
jet power as a function of the accretion rate onto the supermassive black hole, which is the engine of the jet. However, at the best resolution achieved in this work, our simulations 
cannot explicitly resolve accretion onto the central supermassive black hole and the processes that generate the relativistic jet. For this reason, we implement these processes 
following a subgrid approach that bears similarities with those of \cite{2014ApJ...789...54L}, \cite{2017ApJ...847..106L} and \cite{2017ApJ...841..133M}. 

Jet heating is implemented in the same routine that performs radiative cooling. At each hydrodynamical time step the code computes the accretion rate onto the supermassive 
black hole. Since we do not resolve the accretion flow, we assume that the accretion rate is set by the loss of angular momentum of cold gas available in the central regions 
over a characteristic accretion time scale $t_{\rm acc} = 5$ Myr. {This time scale is comparable to the dynamical time at the centre of the BCG}. 
The accretion rate onto the central supermassive black hole is then given by
\begin{equation} \label{eq:accretion}
\dot{M}_{\rm acc} = \frac{M(r<R_{\rm acc}, T<T_{\rm acc})}{t_{\rm acc}},
\end{equation}
where $M(r<R_{\rm acc}, T<T_{\rm acc})$ is the `cold' ($T<T_{\rm acc}$) gas mass present within a sphere of radius $R_{\rm acc}=1.8$ kpc placed at the centre of the computational 
domain. The underlying assumption is that only gas that can cool efficiently will be able to lose angular momentum and fuel the central supermassive black hole. The accretion 
temperature is set to be smaller than the temperature in the initial conditions, but larger than the temperature floor. We chose 
$T_{\rm acc} = 5.0\times10^{5}$ K and $T_{\rm acc} = 1.0\times 10^6$ K for the HR and MR cases, respectively. 

Once the accretion rate has been computed for a given time step $\Delta t$, we remove an amount of mass $\dot{M}_{\rm acc} \Delta t$ from a sphere of radius $R_{\rm acc}$. 
The density of the $i$-th cell within the accretion sphere is updated at each time step:
\begin{equation} \label{eq:acc_scheme}
\rho_{i, {\rm new}}= \rho_{i, {\rm old}} \max\left[0.1, \left(1-\frac{\dot{M}_{\rm acc}\Delta t}{M_{\rm old}} \right)\right],
\end{equation}
where $M_{\rm old}$ is the total gas mass within the accretion sphere before accretion is performed. Equation~\ref{eq:acc_scheme} prevents the accretions scheme from removing 
more than $90\%$ of the mass in a cell within a single time step, which helps prevent the development of cells with negative density. The mass that is removed by accretion 
is not stored, since we do not track the evolution of the mass of the central supermassive black hole. 

Under the assumption that a fraction $\epsilon$ of the accreted rest mass energy is converted into jet power, the latter is computed by:
\begin{equation}\label{eq:jetpow}
\dot{E}_{\rm jet} = \epsilon \dot{M}_{\rm acc} c^2. 
\end{equation}
At each time step $\Delta t$, the jet power is computed using equation~\ref{eq:jetpow} and an amount of energy 
$E_{\rm jet} = \dot{E}_{\rm jet}\Delta t$ is directly injected and equally redistributed in two discs of thickness one cell and radius $0.5\times R_{\rm acc}$, which act as 
jet launching platforms. These discs lie on a plane 
orthogonal to the $z$-axis of the box and are placed 2 cells above and below the box centre, respectively. A fraction $f_{\rm kin} E_{\rm jet}$ is injected in kinetic form, 
whereas a fraction $(1-f_{\rm kin})E_{\rm jet}$ is injected in thermal form. 

We implement jets with fixed velocity $v_{\rm jet} = 10^4$ km/s, but we redistribute the mass in the jet-launching discs using a Gaussian profile. For each cell in a disc we define 
a Gaussian weight:
\begin{equation}
w_i = \frac{\exp\left[-\frac{x_i^2+y_i^2}{2R_{\rm jet}^2}\right]}{\sum\limits_{j, {\rm disc}} \exp\left[-\frac{x_j^2+y_j^2}{2R_{\rm jet}^2}\right]} 
\end{equation}
where the Gaussian smoothing radius is set to $R_{\rm jet} = 0.375 \times R_{\rm acc}$. The kinetic energy injected in each cell of a jet-launching disc is given by
\begin{align} \label{eq:ekin}
\nonumber E_{{\rm kin}, i} & = \frac{1}{2}f_{\rm kin}E_{\rm jet} w_i  \\ 
& = \frac{1}{4}\rho_{{\rm jet}, i} v_{\rm jet}^2\Delta x^3, 
\end{align}
where the factor 1/2 on the r.h.s. on the first line is introduced because the energy associated with the jets has to be redistributed between two discs. Equation~\ref{eq:ekin} 
effectively defines $\rho_{{\rm jet}, i}$, the jet density associated with the $i$-th cell. Each cell in the jet-launching disc receives specific momentum:
\begin{equation}
p_{{\rm jet,} i} = \rho_{{\rm jet}, i}v_{\rm jet} 
\end{equation}
When $f_{\rm kin} < 1$, thermal energy is also injected in each cell belonging to a jet-launching disc using a similar scheme:
\begin{equation}
E_{{\rm th}, i} = \frac{1}{2}(1-f_{\rm kin})E_{\rm jet} w_i.
\end{equation}
When $f_{\rm kin} = 1$ only kinetic energy is injected (purely kinetic jet) and the thermal energy is left to the pre-injection value. This prevents the jet 
from having a formally infinite Mach number. 

{Our chosen hydro jet velocity $v_{\rm jet} = 10^4$ km/s is sub-relativistic, whereas real jets are relativistic. Our choice allows us to perform a direct comparison 
with previous work on hydro jets which were also assumed to be sub-relativistic. More practically, we also tried setting up jets with velocity $10^5$ km/s, but we found the time step to be 
prohibitively small. Since the jet power is set independently of the jet velocity, launching a sub-grid jet at $10^4$ km/s is equivalent to making assumptions on the propagation of the jet 
at un-resolved scales. If in reality a significant fraction of the kinetic energy is quickly thermalized, e.g. via strong shocks near the jet launching region, then this may be approximated by 
our mixed thermal/kinetic models. It is beyond the scope of this work to determine whether relativistic effects may have important consequences on large scales. These assumptions need to 
be explicitly verified in future work.}

{Our fiducial choices for the parameters regulating jet heating are $\epsilon = 0.01$ and $f_{\rm kin} = 1.0$ (fully kinetic jet), however we also explore models 
with lower $\epsilon$ and with hybrid kinetic and thermal feedback ($f_{\rm kin} = 0.5$). Extensive tests of the effects of the variation of $\epsilon$ and $f_{\rm kin}$ at the resolution reached by our 
simulations have been performed in the literature \citep[see ][]{2017ApJ...841..133M}. The efficiency $\epsilon$ essentially regulates the duty cycle of the jet, but has little influence on 
the magnitude of the jet power. The effect on the duty cycle can be understood by noting that the jet power is set by the accretion rate at all times. Quickly after the jet is turned on, the accretion rate 
settles to a value for which the jet power is sufficient to offset the local cooling rate in the centre. As soon as the jet has enough power, it locally heats and displaces gas, the accretion rate 
decreases, followed by the decrease of the jet power. The efficiency therefore determines how quickly the threshold jet power will be reached; the higher $\epsilon$ the faster that will happen. 

Our fiducial value of $\epsilon = 0.01$ is somewhat higher compared to the values used by other authors using similar galaxy cluster simulations 
\citep{2014ApJ...789...54L, 2015ApJ...811...73L, 2016ApJ...829...90Y, 2017ApJ...847..106L, 2017ApJ...841..133M}. \cite{2015ApJ...811...73L} performed jet simulations which also included star formation and showed that a range of $\epsilon$ can achieve self-regulation, but that $\epsilon \gtrsim 0.01$ is required to avoid overproducing the stellar mass of the BCG. 
Previous work adopted a lower $T_{\rm floor}$ and $T_{\rm acc}$, which modify the net amount of matter accreted by the central supermassive black hole, and consequently the jet power triggered by the cooling flow. Therefore, it is not surprising that our simulations achieve similar results with a different $\epsilon$. The appropriate value for $\epsilon$ can be constrained by going to much higher resolution and by more appropriately resolving the phase structure of the ICM \citep[e.g., ][]{2014ApJ...789...54L}, which is unfeasible at the resolution we achieve. For the purposes of our simulations, $\epsilon$ must be treated as a subgrid parameter that allows to achieve realistic values for the jet power given the cooling flow, which is the most important aspect for studies of jet heating in cluster cores. }
 
To enhance the interaction of the jet material with the ICM we also implement jet precession: the velocity vector of the jet forms an angle $\theta_{\rm prec}$ with the $z$-axis
and precesses around the latter over a period of $t_{\rm prec} = 10$ Myr. We adopt a fiducial value for $\theta_{\rm jet} = 15^\circ$ for the precession angle, but we also performed tests with 
$\theta_{\rm jet}= 30^\circ$ that provided qualitatively similar results. For this reason, tests with $\theta_{\rm jet}= 30^\circ$ will be omitted in our analysis below. 
Conversely, using smaller values of $\theta_{\rm jet}$ leads to jets that 
pierce through the ICM without producing considerable heating, as shown in past literature \citep{2004MNRAS.348.1105O, 2007MNRAS.376.1547C}. 

\begin{table*}
\centering
{\bfseries Catalog of Simulations}
\resizebox{\linewidth}{!}{
\begin{tabular}{llcccccccc}
\hline
Name & Jet Injection &$\Delta x$ [pc]& $T_{\rm floor}$ [K] & $T_{\rm acc}$ [K] & $R_{\rm acc} $ [kpc] & $\epsilon$ & $f_{\rm kin}$ & Riemann Solver and Integrator \\
\hline
LR\_JET\_HLLC  & Purely Kinetic & 1562 & $2.0\times 10^4$ & $5.0\times 10^5$ & 3.2 & 0.001 & 1.0 & HLLC+CTU \\
LR\_MIXED\_HLLC & Mixed Thermal/Kinetic & 1562 & $2.0\times 10^4$ & $5.0\times 10^5$ & 3.2 & 0.001 & 0.5 & HLLC+CTU \\
MR\_COOL\_HLLC & Cooling Only, No Jet & 390 & $5.0\times 10^5$ & n.a. & n.a. & n.a. & n.a. & HLLC+CTU \\
MR\_JET\_HLLE & Purely Kinetic & 390 & $5.0\times 10^5$ & $1.0\times 10^6$ & 1.8 & 0.010 & 1.0 & HLLE+VL \\
MR\_JET\_HLLC & Purely Kinetic & 390 & $5.0\times 10^5$ & $1.0\times 10^6$ & 1.8 & 0.010 & 1.0 & HLLC+CTU \\
MR\_MIXED\_HLLE & Mixed Thermal/Kinetic & 390 & $5.0\times 10^5$ & $1.0\times 10^6$ & 1.8 & 0.010 & 0.5 & HLLE+VL \\
MR\_MIXED\_HLLC & Mixed Thermal/Kinetic & 390 & $5.0\times 10^5$ & $1.0\times 10^6$ & 1.8 & 0.010 & 0.5 & HLLC+CTU \\
HR\_COOL\_HLLC & Cooling Only, No Jet & 195 & $2.5\times 10^5$ & n.a. & n.a. & n.a. & n.a. & HLLC+CTU \\
HR\_JET\_HLLC & Purely Kinetic & 195 & $2.5\times 10^5$ & $5.0\times 10^5$ & 1.8 & 0.010 & 1.0 & HLLC+CTU \\
HR\_MIXED\_HLLC & Mixed Thermal/Kinetic & 195 & $2.5\times 10^5$ & $5.0\times 10^5$ & 1.8 & 0.010 & 0.5 & HLLC+CTU \\
\hline
\end{tabular}
}
\caption{ List of simulations and their parameters. Column 1: simulation name. Column 2: type of jet injection scheme. Column 3: $\Delta x$ is the cell size at the highest level of refinement. Column 4: $T_{\rm floor}$ is the temperature floor. Column 5: $T_{\rm acc}$ is the temperature below which gas is accreted onto the central supermassive black hole. Column 6: $R_{\rm acc} $ is the radius of the spherical region within which the black hole accretion rate is computed. Column 7: $\epsilon$ is the jet efficiency. Column 8: $f_{\rm kin}$ is the fraction of kinetic energy injected by the jet. Column 9: Riemann solver used for the simulation and numerical integrator used for the simulation.}\label{tab:list_sims}
\end{table*}

\section{Diagnostics of Heating and Cooling Rates}

We use a series of diagnostics of the heating and cooling rates as a function of location and time. 
In summary, we estimate:
\begin{itemize}
\item The total cooling rate in each cell of the simulations. 
\item The total heating rate from in each cell of the simulations. 
\item {The kinetic power associated with radial outflowing gas motions}.
\item {An estimate of the turbulent kinetic energy dissipation rate, i.e. an upper limit to the heating rate provided by turbulence in different regions of the cluster. }
\item An estimate of the heating rate from weak shocks. 
\item {An estimate of the power associated with adiabatic compression and expansion (i.e. `$PdV$ power'). }
\end{itemize}
The following subsections describe the methods used to measure each quantity in the list.

\subsection{Total Heating and Cooling Rates}\label{sec:tot_heat}

In our implementation, the cooling rate is stored as an additional variable for each cell and is updated at each time step. To estimate the total heating rate in each cell, 
we implement a version of the scheme developed by \cite{2015MNRAS.454.1848R}, which was originally developed to estimate heating in 
general relativistic magnetohydrodynamics simulations of accreting black holes. We briefly review the main features of this scheme and defer to 
Appendix~\ref{appendix:A} and to \cite{2015MNRAS.454.1848R} for details. 

The heating estimator is based on the fact that conservative codes like {\sc athena} conserve total energy to machine precision, but effectively introduce numerical viscosity, which 
results in dissipation of kinetic energy on small scale, i.e. acts as a source of heating. 
The heating estimator measures the truncation error in the energy equation at each time step and translates it into a heating rate. This estimate 
can be done cell-by-cell. However in cells which experience very small heating, the truncation error can be negative, resulting in negative heating rates. These values are 
not physical if taken at face value. This method only provides physically reliable heating rates if they are averaged over a sufficiently large length or time, as verified  
by \cite{2015MNRAS.454.1848R}, who performed a wide range of tests. 

After implementing the heating estimator in {\sc athena}, we perform a series of tests to assess its accuracy (see Appendix~\ref{appendix:B}). We conclude that the method 
is able to yield heating rates with $\sim 5\%$ accuracy.

\subsection{Kinetic Power and Turbulent Heating Rate }

{Jets accelerate the ICM which may lead to radial motions. The kinetic energy associated with these motions may be a significant fraction of the jet power, especially near the jet-launching region. For this reason, we measure the kinetic power associated with radial motions as:
\begin{equation}
 \dot{E}_{\rm kin,rad} = \oint_S \frac{1}{2}\rho_{\rm gas} v_{\rm rad}^2 \Theta(v_{\rm rad}) \vec{v}_{\rm rad}\cdot d\vec{S'},
\end{equation}
where $S$ is the boundary surface of a sphere of radius $R$, $\rho_{\rm gas}$ is the gas density, $\vec{v}_{\rm rad}$ is the radial component of the gas velocity, $\Theta(v_{\rm rad})$ is a step function that selects only regions where the gas flows away from the centre ($v_{\rm rad}>0$).
}

Injection of a large quantity of momentum by the jets results in significant acceleration of gas, wave driving, inflation of low density cavities that may buoyantly rise and drive 
turbulent motions. To estimate the energy dissipation rate associated with turbulent motions we perform a spectral analysis of velocity fluctuations with respect to the radial motion of the gas. 
Only a fraction of such fluctuations can be attributed to turbulence, therefore they can only be used to infer an upper limit to the turbulent heating rate. {Furthermore, velocity fluctuations may also mix hot gas with colder gas, which is also a source of heating. For these reasons, the `turbulent heating rate' estimated below is an upper limit of the heating rate provided by both 
a turbulent cascade from large scales down to the grid scale and gas mixing happening on large scales.}

We focus on velocity fluctuations relatively to a background which may have net radial velocity (spherically symmetric cooling flow). We define the velocity fluctuations as:
\begin{equation}
 \delta \vec{v}(x) = \vec{v}(x)-\vec{v}_{\rm rad}(x),
\end{equation}
where $\vec{v}_{\rm rad}(x)$ is the radial velocity at radius $x$. $\vec{v}_{\rm rad}(x)$ is estimated by averaging the velocity field in radial shells that do not contain the jet material ($45^\circ < \vartheta < 135^\circ$).
Let $\delta \vec{v}_{\vec{k}}$ be the (discrete) Fourier transform of the velocity fluctuation field, then its power spectrum can be written as:
\begin{equation}
 \sigma^2(k) = \sum_{|\vec{k}|=k} |\delta \vec{v}_{\vec{k}}|^2.
\end{equation}
After computing the velocity fluctuation power spectrum, we define an effective driving scale for turbulent motions:
\begin{equation}
 k_{\rm drive} = \frac{\sum_{|\vec{k}|=k} k|\delta \vec{v}_{\vec{k}}|^2}{ \sum_{|\vec{k}|=k} |\delta \vec{v}_{\vec{k}}|^2 },
\end{equation}
i.e., the effective driving scale is achieved by a weighted average of the $k$-s associated with each mode with the weights given by the power spectrum. Finally, we estimate the kinetic energy dissipation rate as:
\begin{equation}\label{eq:turb}
 \dot{E}_{\rm kin,turb} = \frac{1}{2}\bar{\rho}[\sigma^2(k_{\rm drive})]^{3/2}k_{\rm drive},
\end{equation}
where $\bar{\rho}$ is the average density in the region where the power spectrum is computed, $\sigma^2(k_{\rm drive})$ is the velocity
dispersion of the velocity fluctuations with respect to the radial velocity field estimated at the effective driving scale $k_{\rm drive}$. Equation~\ref{eq:turb} is derived by assuming that the kinetic energy $E_{\rm kin,turb}\sim 0.5\bar{\rho}\sigma^2$ associated with turbulent eddies of size $L\sim1/k_{\rm drive}$ will cascade cascade down to eddies of smaller size in approximately one eddy turnover time $t_{\rm turn} = (k_{\rm drive} \sigma)^{-1}$.

\subsection{Shock Heating Rate}

The ICM can also be heated by weak shocks. We adopt an approximate estimator of shock heating which is applied in post-processing and which uses theoretical arguments similar to those used by \cite{2016ApJ...829...90Y}. Shocks are identified in our simulations by measuring entropy, pressure and density jumps. 

The pressure jump across a shock can be expressed as:
\begin{equation}\label{eq:dpoverp}
\frac{\Delta P}{P} = \frac{P_2-P_1}{P_1} = \frac{2\gamma}{\gamma+1}y,
\end{equation}
where $P_1$ and $P_2$ are the pre-shock and post-shock pressures, respectively and $y$ is a dimensionless parameter related to the shock Mach number $M_{\rm s}$:
\begin{equation}\label{eq:mach}
y = \frac{\rho_1v_{\rm s}^2}{\gamma P_1} -1 = M_{\rm s}^2 - 1,
\end{equation}
where $\rho_1$ is the pre-shock density. The density jump across the shock is given by:
\begin{equation}\label{eq:rhojump}
\delta = \frac{\rho_2}{\rho_1} = \frac{(\gamma+1)(y+1)}{2+(\gamma-1)(y+1)}.
\end{equation}
Finally, the entropy jump across a weak shock is given by:
\begin{equation}
ds \approx \frac {2\gamma k_{\rm B}}{3(\gamma+1)^2\mu m_{\rm H}}y^3.
\end{equation}
We measure density, pressure and entropy jumps in all cells in the computational volume by computing differences between consecutive snapshots. Snapshots are saved every $\Delta t = 5$ Myr, which is the effective time resolution of our shock heating estimator. A cell is flagged as shock heated only when the density jump is $\rho_2/\rho_1 > 0.9 \delta$, $ds > 0$ and $\Delta P/P > 0.2$. The first condition makes sure that we are measuring density jumps that are representative of shocked regions, i.e. where the density jump is similar to that given by the shock jump condition of equation~\ref{eq:rhojump}. The second condition makes sure that there is local heating, which is associated with an entropy jump. The third condition makes sure that we detect shocks with Mach number larger than a given threshold. In fact, once $\Delta P/P$ is measured, $y$ is calculated from equation~\ref{eq:dpoverp} and is associated with a Mach number using equation~\ref{eq:mach}. Selecting $\Delta P/P > 0.2$ corresponds $M_{\rm s} > 1.08$. By varying the threshold for the pressure jump, we empirically determine that shock heating comes mostly from shocks with $M_{\rm s}\gtrsim 1.1$. 

Finally, once a cell has been flagged as shock heated, the shock heating rate is then estimated as:
\begin{equation}
\dot{E}_{\rm shock} = \frac{\rho_1T_1ds}{\Delta t},
\end{equation}
where $T_1$ is the pre-shock temperature. 

In principle, if the time resolution is too coarse, the shock heating estimator described in this subsection may not take into account the contributions of shocks with velocity $v_{\rm s} > \Delta x/\Delta t$. We test the robustness of the shock heating estimator by varying the time resolution from $\Delta t = 5$ Myr to $\Delta t = 1$ Myr, but we do not find significant differences in the estimate of the shock heating rate. The result of this test confirms that the contribution of strong shocks to the heating rate is negligible compared to that from weak shocks in the regions away from the jet axis. The situation is more complicated within the jet cone, where highly supersonic material is present and where heating from strong shocks may provide a large contribution. In this region, even a time resolution of 1 Myr might not be enough to capture most of the heating from shocks, implying that our shock heating estimator is only able to provide a lower limit to the total shock heating rate in the jet cone.

{As a final caveat, we should stress that weak shocks can also be generated by turbulence. The effect will be more prominent for supersonic turbulence and less prominent for transonic and subsonic turbulence. For this reason, it is possible that the shock heating rate estimator of this subsection and the turbulent heating rate of the previous subsection probe the same physical processes in some circumstances. In principle, the two heating rate estimator should differ significantly from each other in regions where turbulence is subsonic, whereas they should be similar to each other in regions where turbulence is transonic/supersonic. We discuss this issue in more detail in the results section. }

\subsection{Power from adiabatic compression and expansion}

{Despite not contributing to changes of the gas entropy along the flow, forces that cause adiabatic compression/expansion also cause changes in the internal energy of the ICM by doing $PdV$ work. For this reason, adiabatic compression/expansion may significantly influence the thermodynamics of the ICM. We estimate the $PdV$ power following \cite{2016ApJ...829...90Y}, i.e. by calculating the following integral: }
\begin{equation} \label{eq:adirate}
\dot{E}_{\rm PdV} = -\oint_S P\vec{v}\cdot d\vec{S'}+\int_V \vec{v}\cdot\nabla P dV',
\end{equation}
{where $V$ is the volume within which the heating rate is measured and $S$ is its boundary surface. When the luminosity in equation~\ref{eq:adirate} is positive, it corresponds to an increase in the thermal energy of the gas, which may contrast radiative cooling. We note, however, that the cooling flow predicts a $PdV$ power comparable to the cooling rate (Appendix~\ref{appendix:C}), so care must be taken interpreting this diagnostic.}


\section{Results}

\subsection{Fiducial Runs - Fully Kinetic Jet}

\begin{figure*}
\begin{center}
    \includegraphics[width=0.49\textwidth]{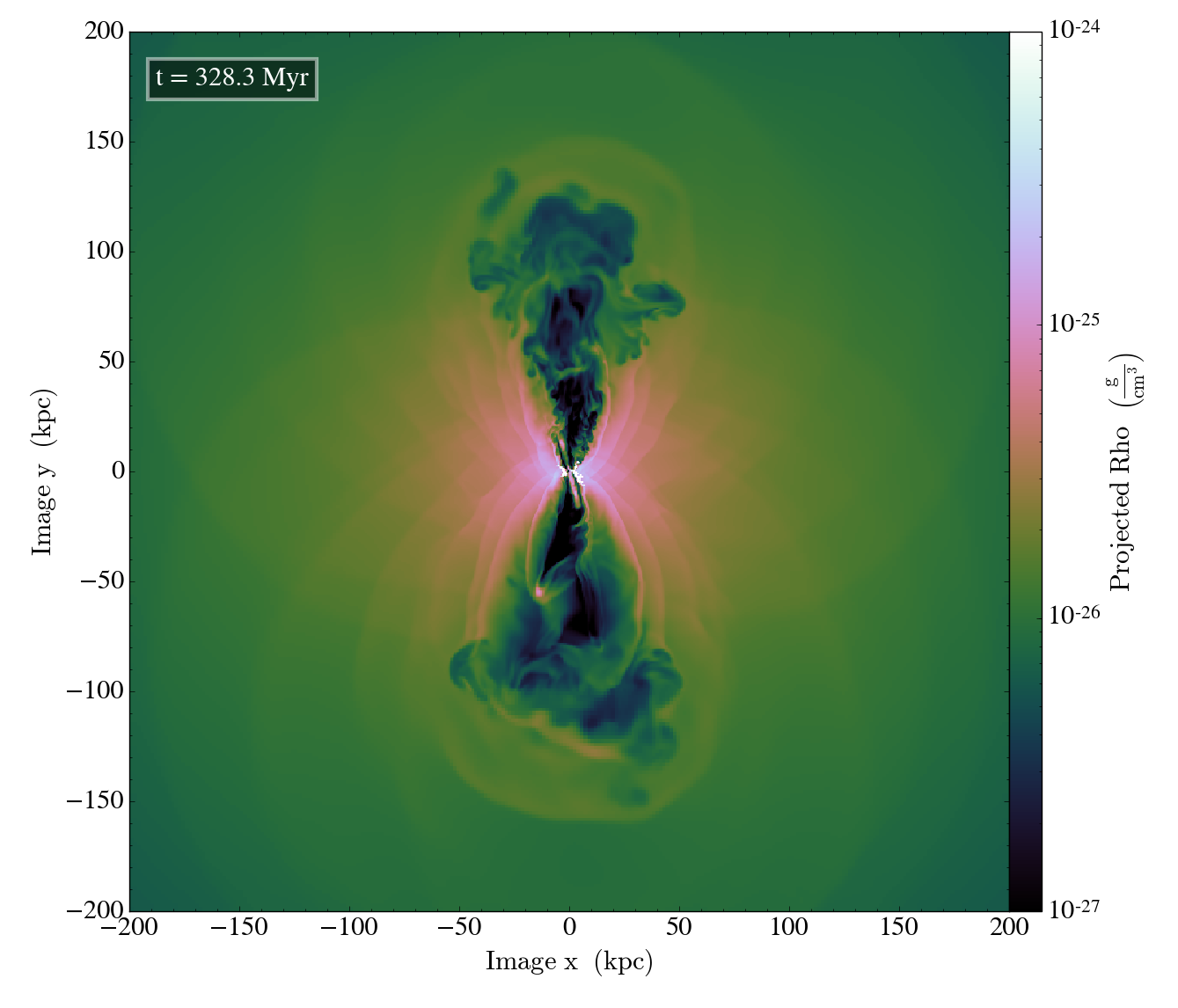}
    \includegraphics[width=0.49\textwidth]{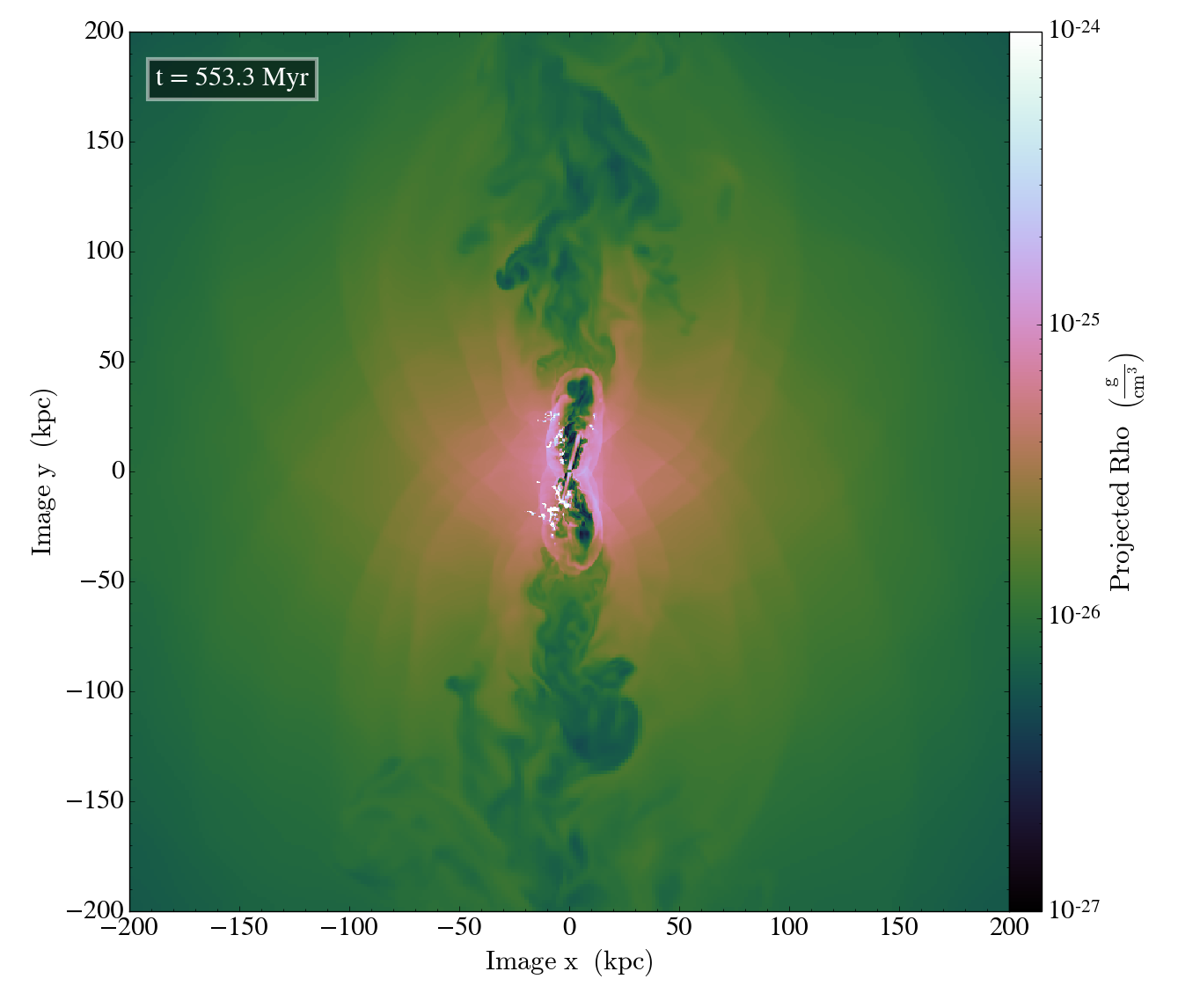}
    \includegraphics[width=0.49\textwidth]{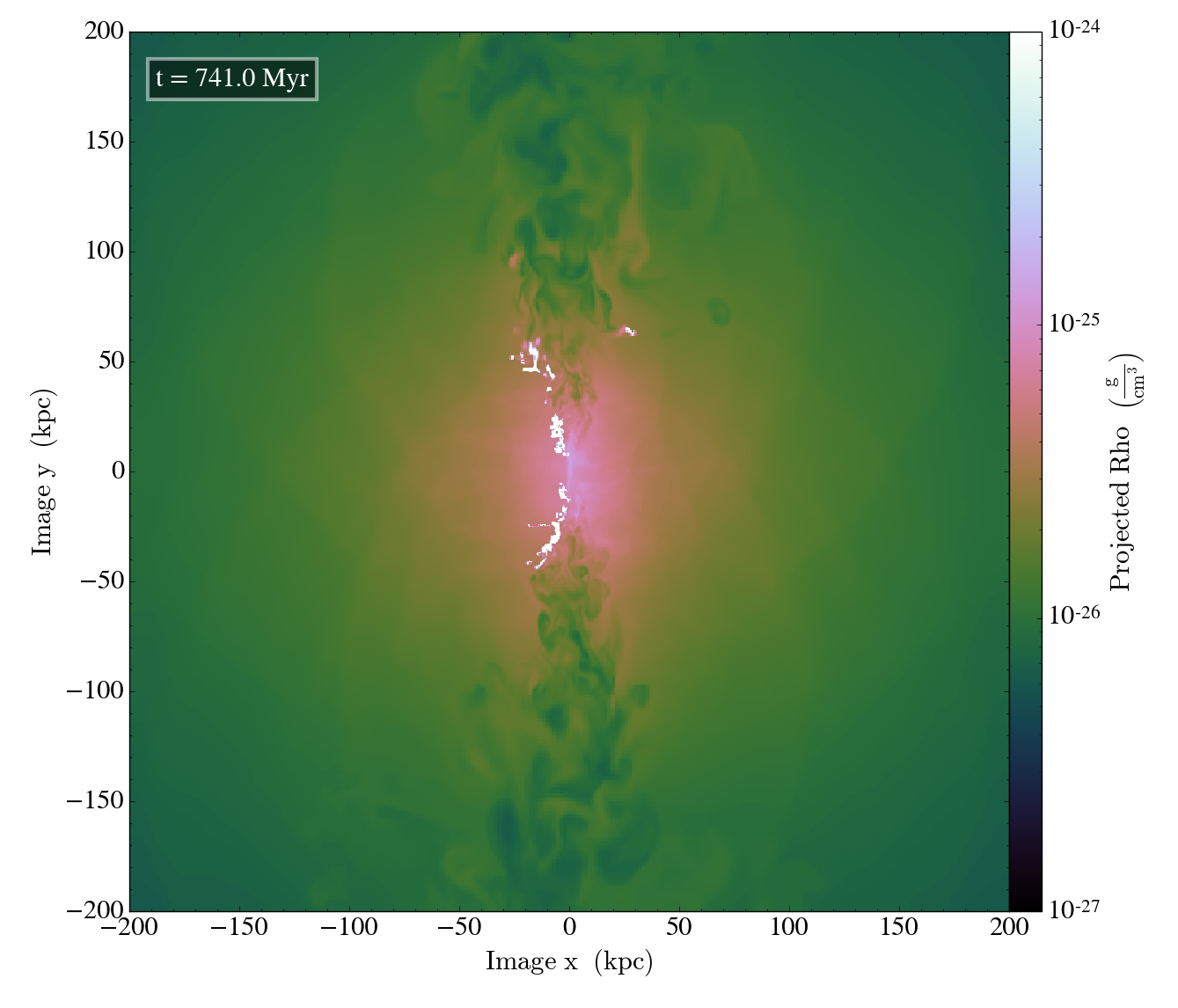}
    \includegraphics[width=0.49\textwidth]{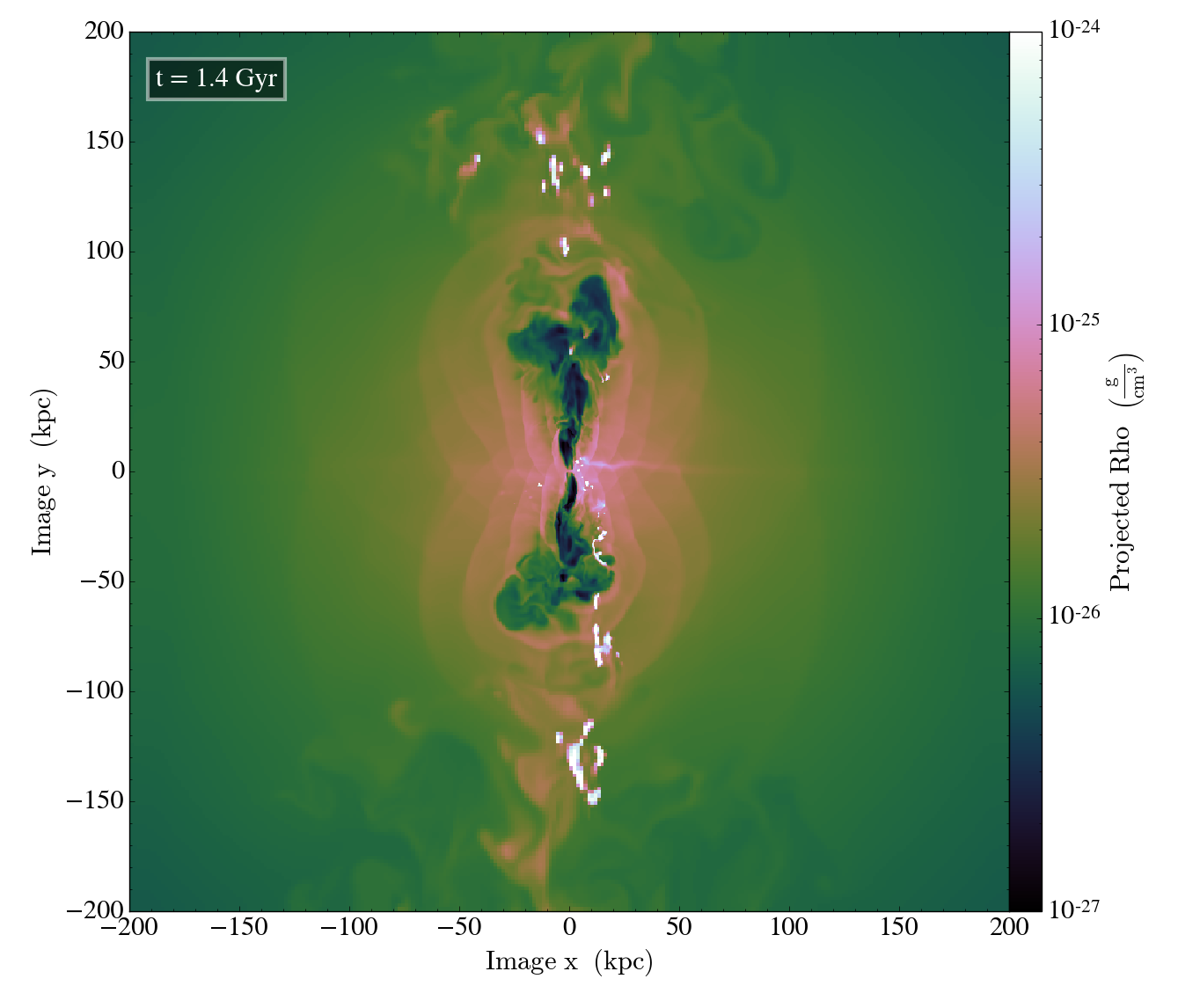}
\end{center}
\caption{ Average density of the gas in thin slices at four different times for the ${\rm HR\_JET\_HLLC}$ simulation which adopts a purely kinetic jet and has spatial resolution $\sim 200$ pc. The thickness of each slice is 5 kpc. Top-left: jet is on. Top-right: jet recently switched off. Bottom-left: jet is off. Bottom-right: jet is on again. }\label{fig:maps}
\end{figure*}

\begin{figure*}
\begin{center}
    \includegraphics[width=0.49\textwidth]{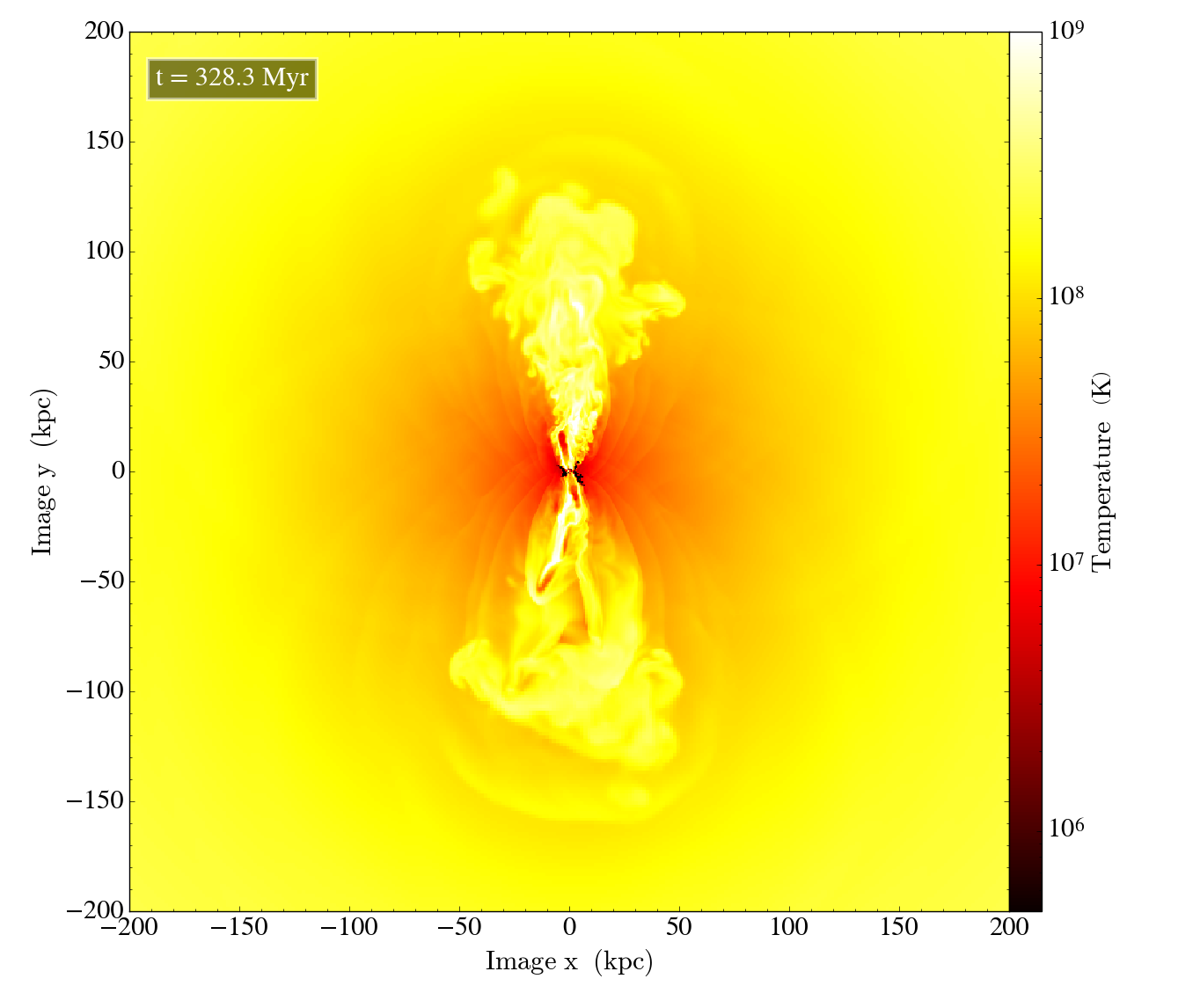}
    \includegraphics[width=0.49\textwidth]{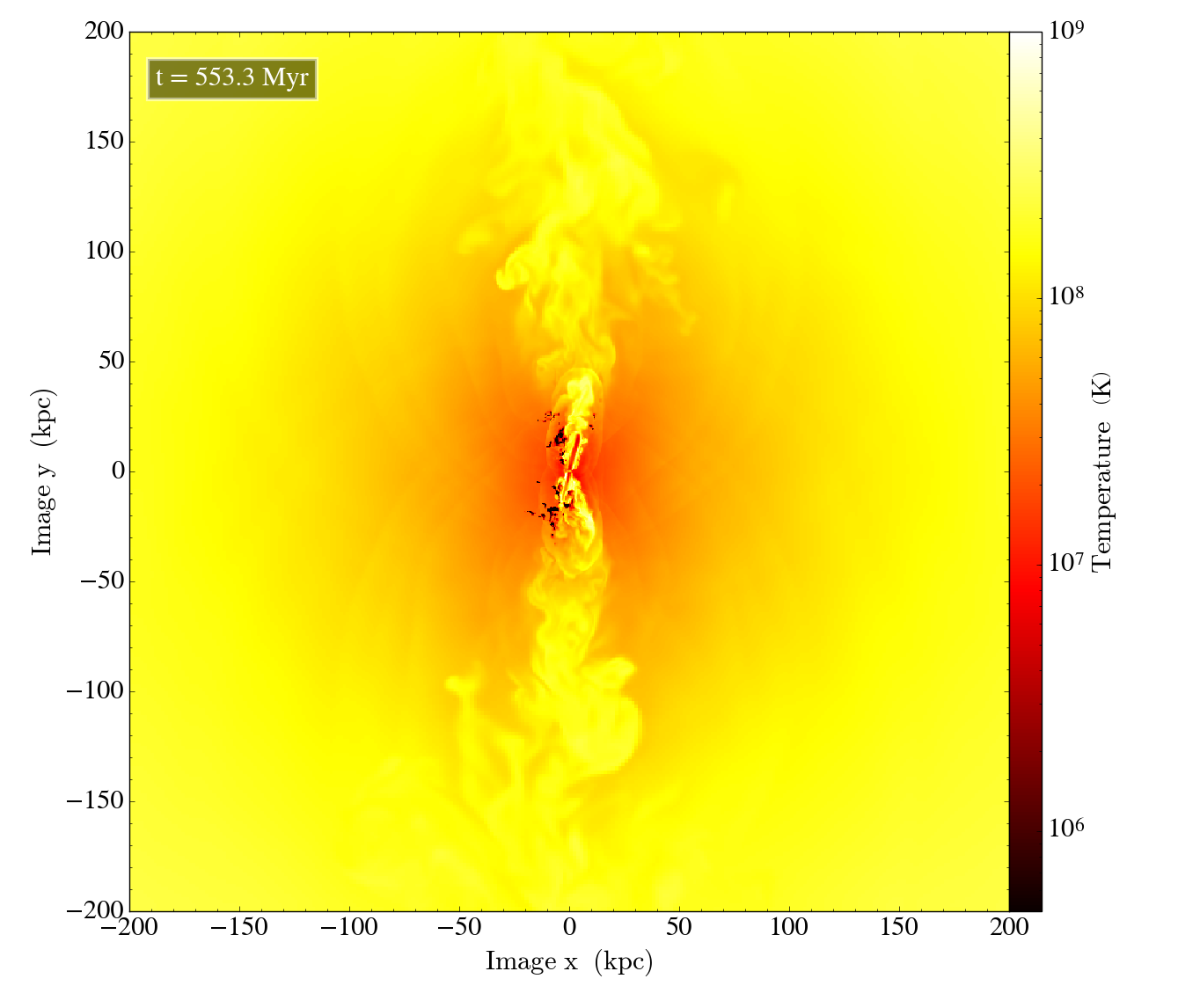}
    \includegraphics[width=0.49\textwidth]{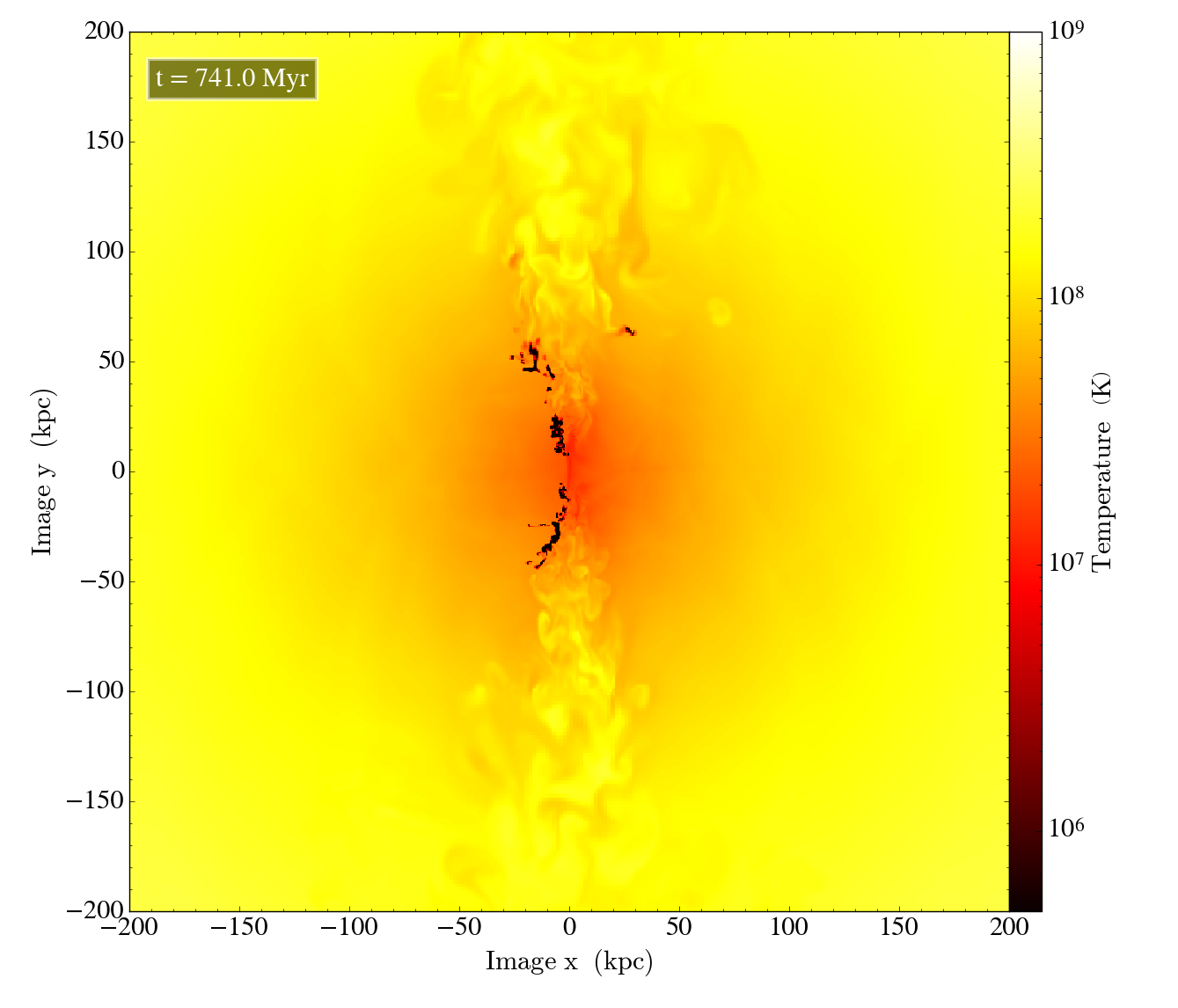}
    \includegraphics[width=0.49\textwidth]{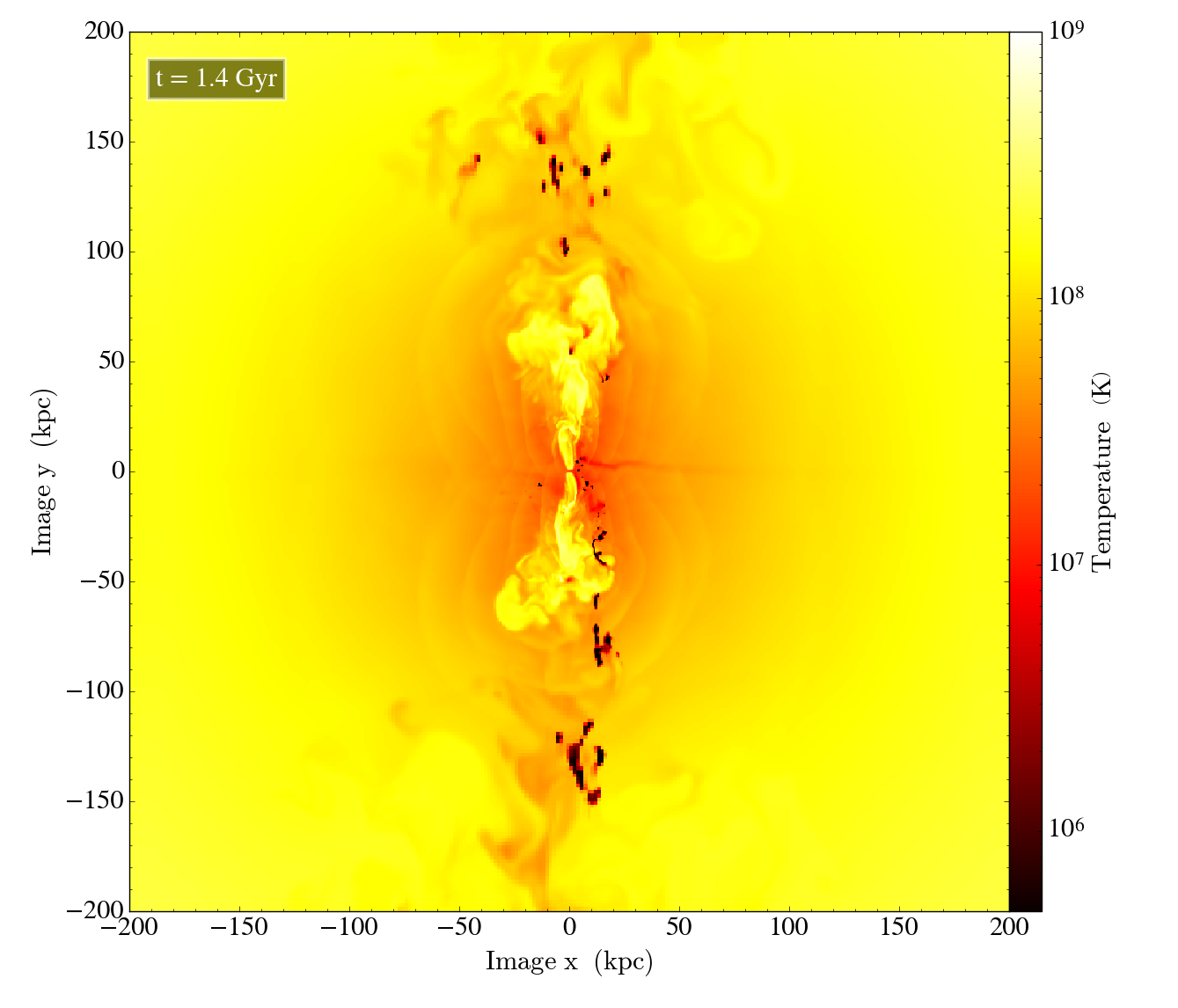}
\end{center}
\caption{ Average temperature of the gas in thin slices at four different times for the ${\rm HR\_JET\_HLLC}$ simulation which adopts a purely kinetic jet and has spatial resolution $\sim 200$ pc. The thickness of each slice is 5 kpc. Top-left: jet is on. Top-right: jet recently switched off. Bottom-left: jet is off. Bottom-right: jet is on again. }\label{fig:mapsT}
\end{figure*}

We consider ${\rm MR\_JET\_HLLC}$ and ${\rm HR\_JET\_HLLC}$ as our fiducial runs because they use the HLLC Riemann solver which captures large density contrasts and multi-phase structures better than HLLE and more diffusive solvers. Figure~\ref{fig:maps} and Figure~\ref{fig:mapsT} show average density and average temperature maps of the gas in the ${\rm HR\_JET\_HLLC}$ simulation in thin slices at four different times. The top-left panel shows a phase in which the jet is on and has inflated low density cavities and lifted significant amount of gas along the jet axes. The top-right panel shows a snapshot in which the jet power has decreased due to reduced gas accretion towards the central region, and in which high density clumps started forming via thermal instability. In the bottom-left panel the jet is completely off and more high-density clumps have formed. The bottom-right panel shows a phase in which the jet has been turned on again and in which some of the clumps have been lifted upwards, whereas others have precipitated towards the centre and fed the central AGN. These figures clearly demonstrate the multi-phase structure of the ICM developed in these simulation, which is a major reason why Riemann solvers that capture contact discontinuities such as HLLC should be favored. 

\begin{figure*}
\begin{center}
    \includegraphics[width=0.49\textwidth]{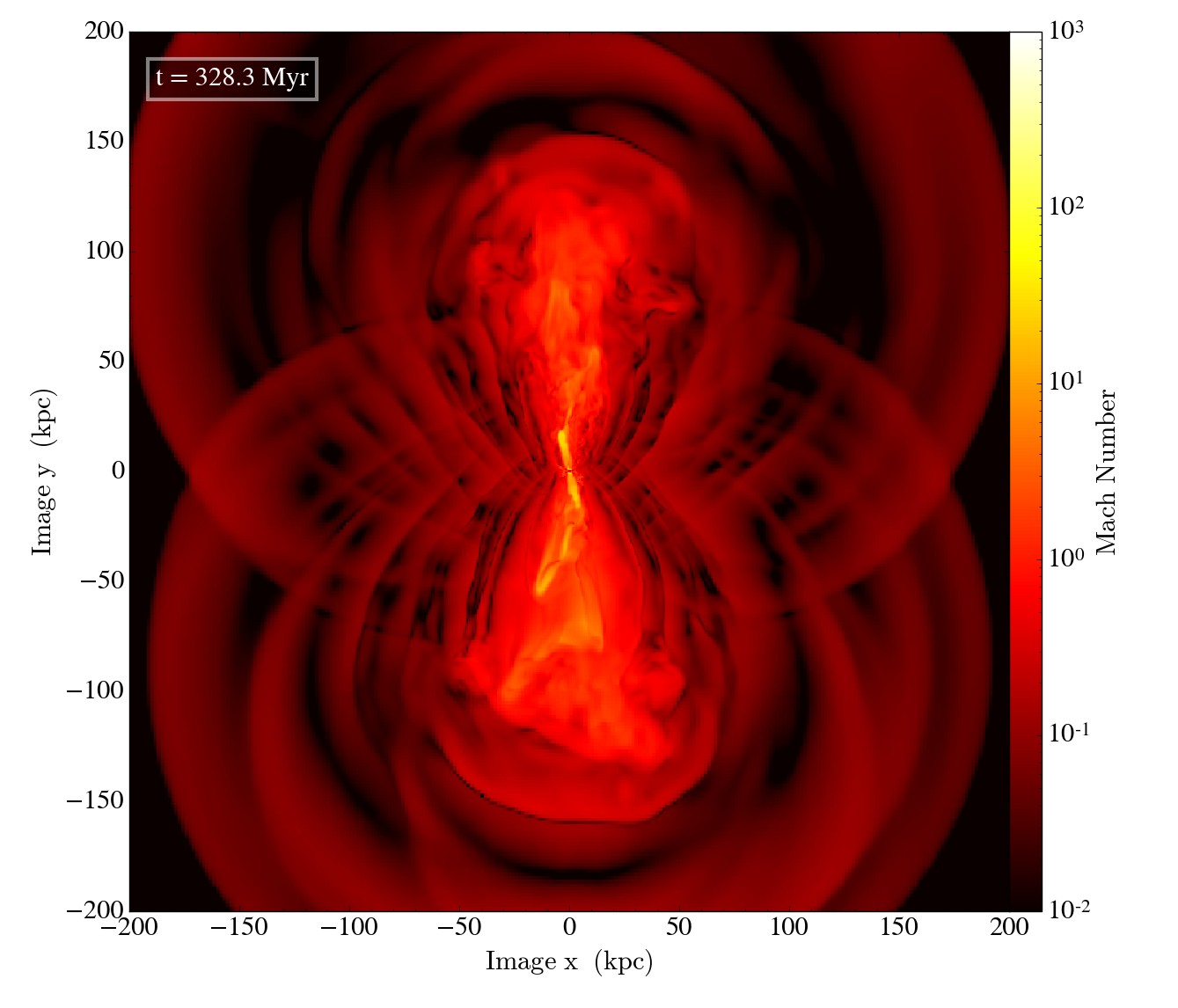}
    \includegraphics[width=0.49\textwidth]{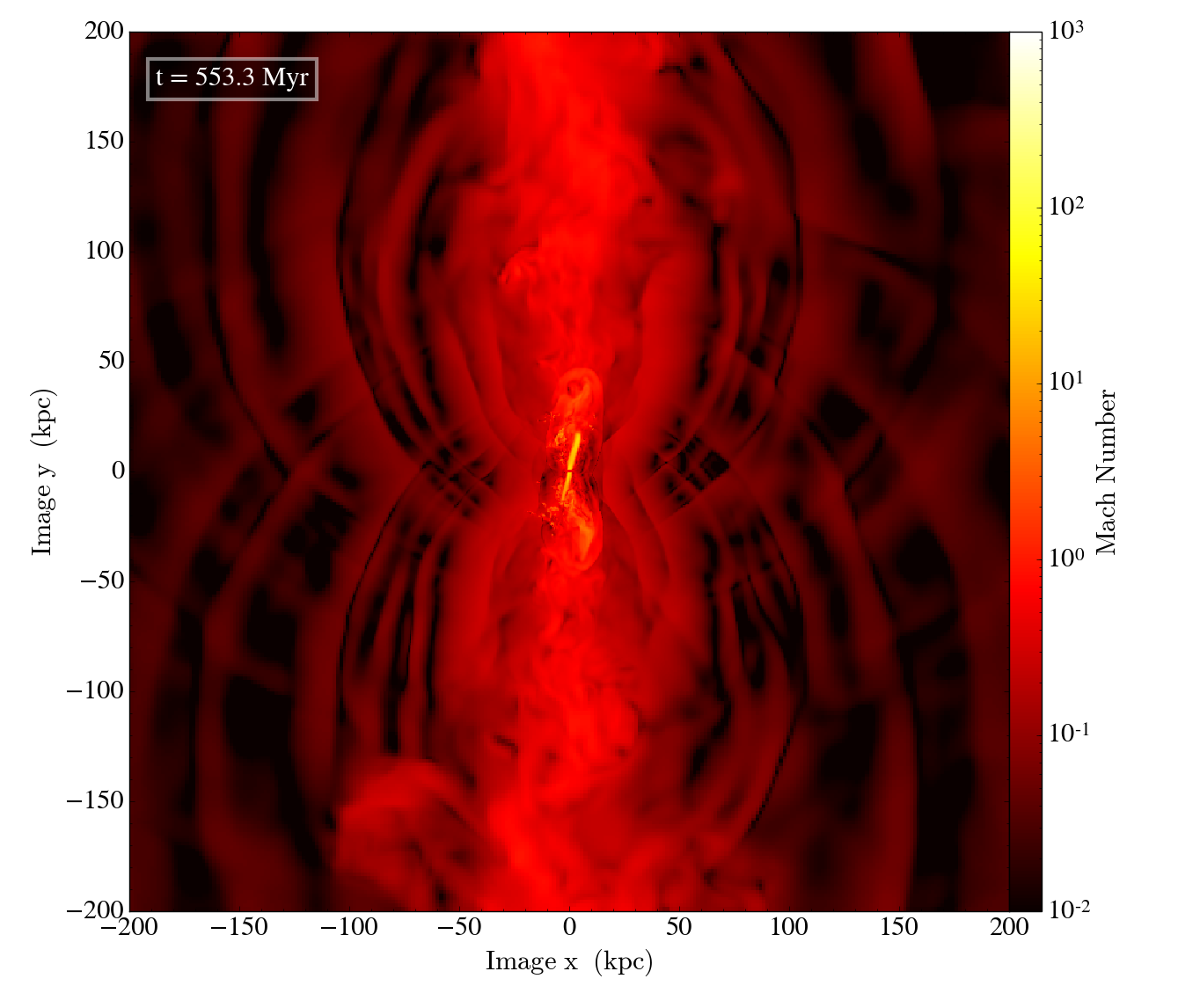}
    \includegraphics[width=0.49\textwidth]{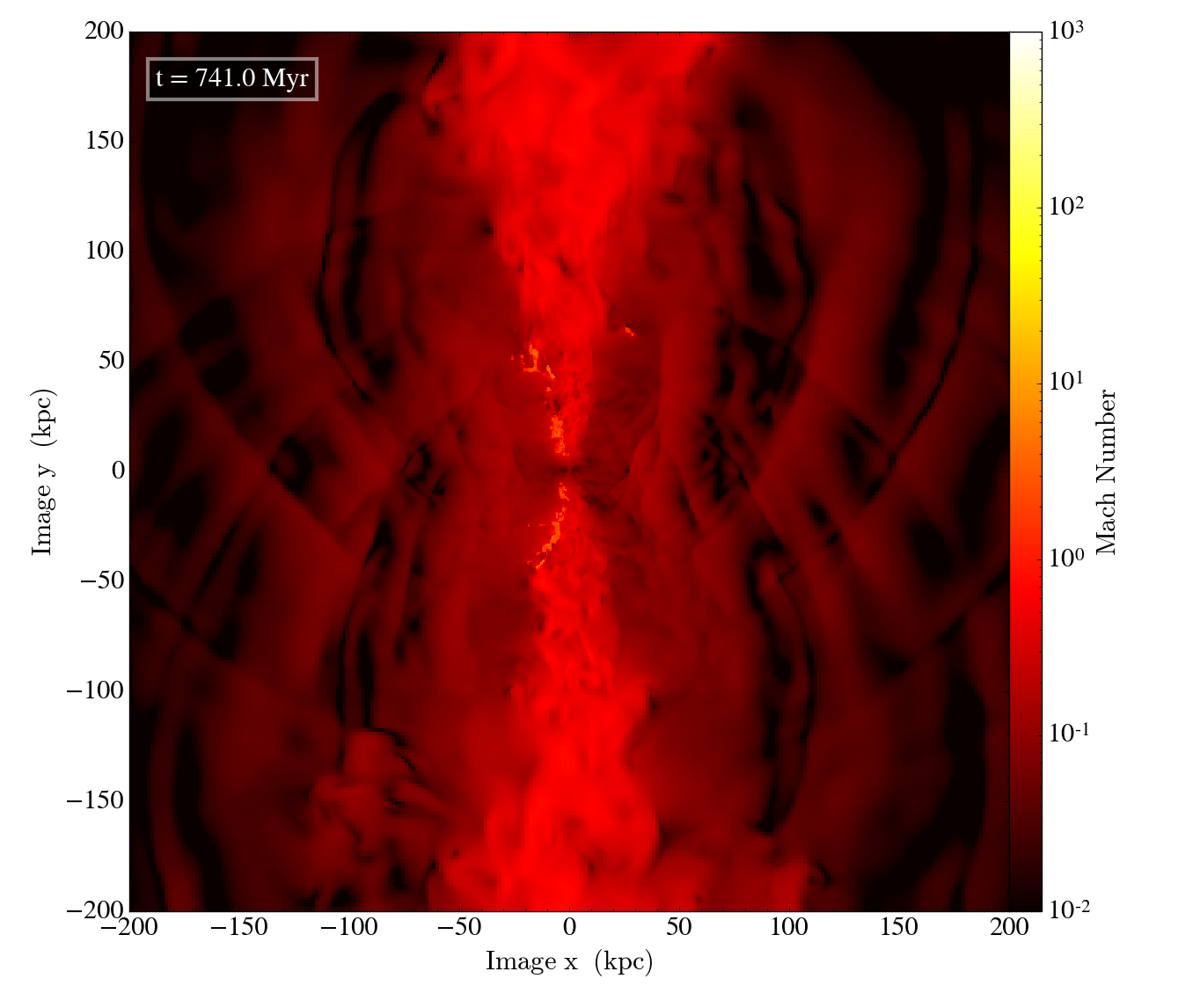}
    \includegraphics[width=0.49\textwidth]{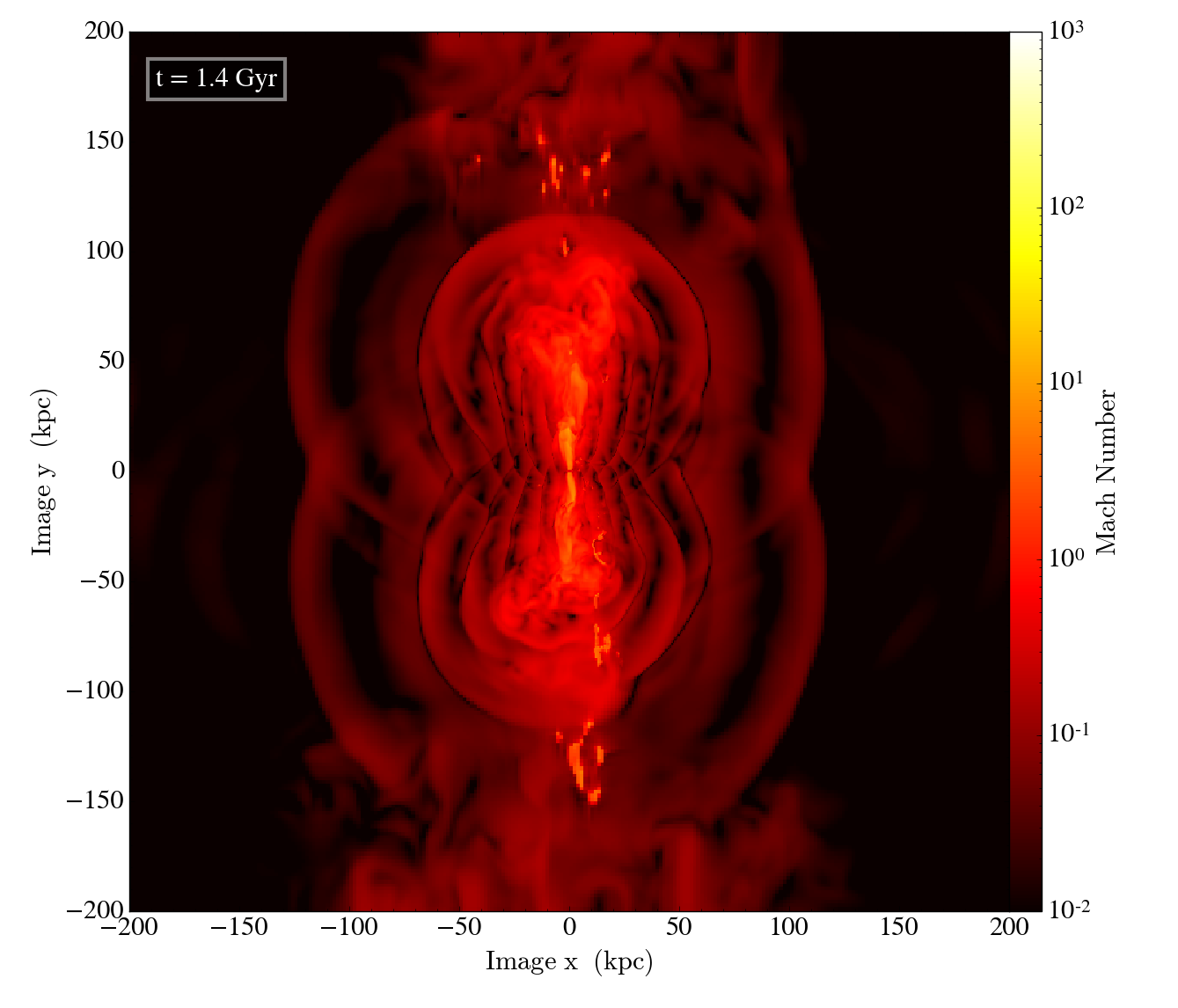}
\end{center}
\caption{ Average Mach number of the gas in thin slices at four different times for the ${\rm HR\_JET\_HLLC}$ simulation which adopts a purely kinetic jet and has spatial resolution $\sim 200$ pc. The thickness of each slice is 5 kpc. Top-left: jet is on. Top-right: jet recently switched off. Bottom-left: jet is off. Bottom-right: jet is on again. }\label{fig:maps_Mach}
\end{figure*}

{Figure~\ref{fig:maps_Mach} shows maps of the Mach number of the flow at the same times shown in Figures~\ref{fig:maps} and \ref{fig:mapsT}. The flow is supersonic only in a small region of the volume, near the injection region of the jets. In the conical regions swept by the precessing jet the flow is mildly transonic or subsonic. Away from the jet cones the flow is typically subsonic. Under these circumstances, turbulent motions can contribute significantly to the generation of shock heating only in specific regions of the jet cone where the flow is transonic.}

\begin{figure*}
\begin{center}
    \includegraphics[width=0.99\textwidth]{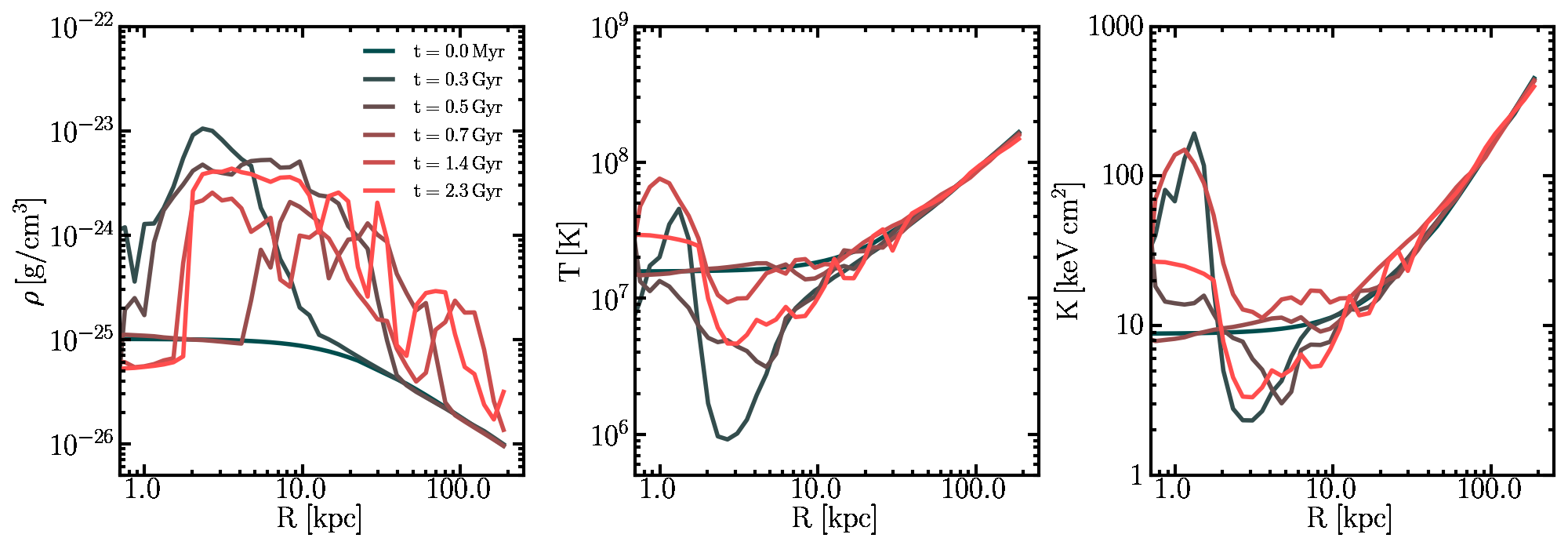}
\end{center}
\caption{ Mass-weighted density (i.e. $<\rho^2>/<\rho>$, left panel), temperature (central panel) and entropy (right panel) profiles at different times in the ${\rm HR\_JET\_HLLC}$ simulation which adopts a purely kinetic jet and has spatial resolution $\sim 200$ pc. }\label{fig:profilesm}
\end{figure*}

\begin{figure*}
\begin{center}
    \includegraphics[width=0.99\textwidth]{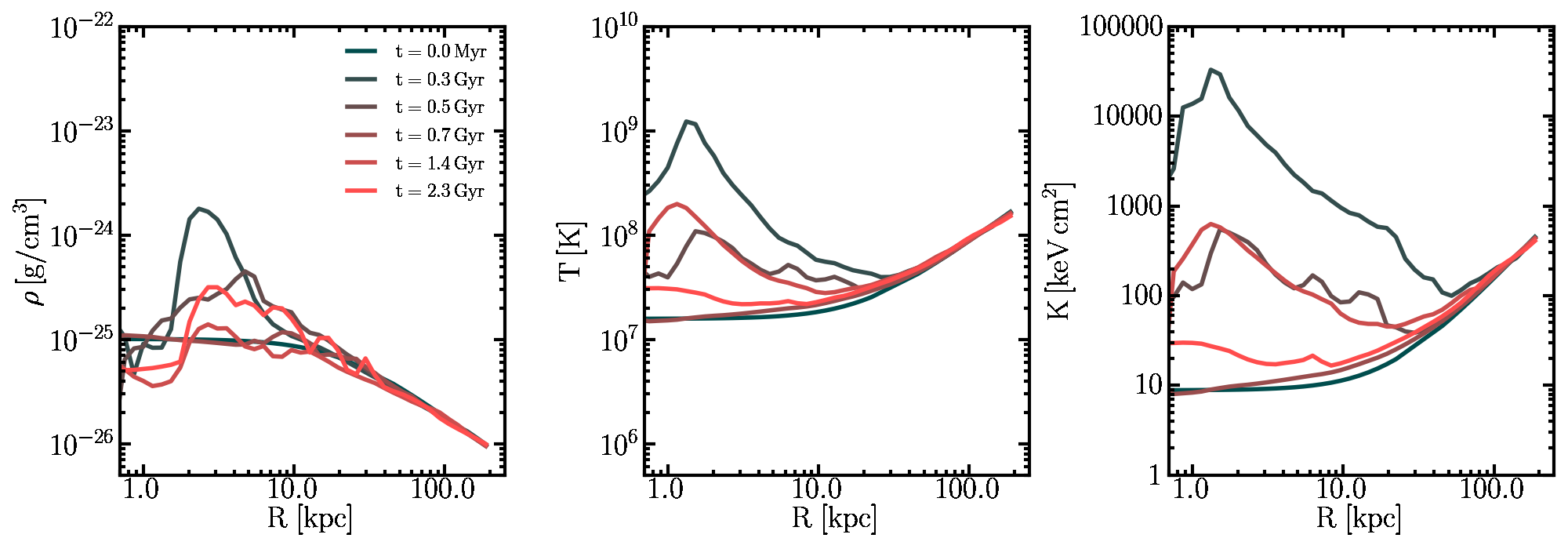}
\end{center}
\caption{ Volume-weighted density (left panel), temperature (central panel) and entropy (right panel) profiles at different times in the ${\rm HR\_JET\_HLLC}$ simulation which adopts a purely kinetic jet and has spatial resolution $\sim 200$ pc. }\label{fig:profilesv}
\end{figure*}

{Figure~\ref{fig:profilesm} and Figure~\ref{fig:profilesv} show the mass-weighted and volume-weighted density, temperature and entropy profiles at different times in the ${\rm HR\_JET\_HLLC}$, respectively. The ${\rm HR\_JET\_HLLC}$ simulation adopts a purely kinetic jet and has spatial resolution $\sim 200$ pc.} The fluid properties at radius $R<5$ kpc are very time variable, because of the jet activity. During the first $0.5$ Gyr the cluster undergoes an initial cooling flow which drives the central density to increase and the temperature and entropy to decrease. The cooling flow is offset by the jet activity once it turns on. At $t>0.5$ Gyr the average density in the central 50 kpc stays higher than in the initial conditions, but the temperature never descends below $10^6$ K. At late times $t>0.7$ Gyr the entropy profile at $R>5 $ kpc reaches a quasi-steady state and a dramatic drop of the central entropy associated with a cooling catastrophe is not observed. 

\begin{figure}
\begin{center}
\includegraphics[width=0.49\textwidth]{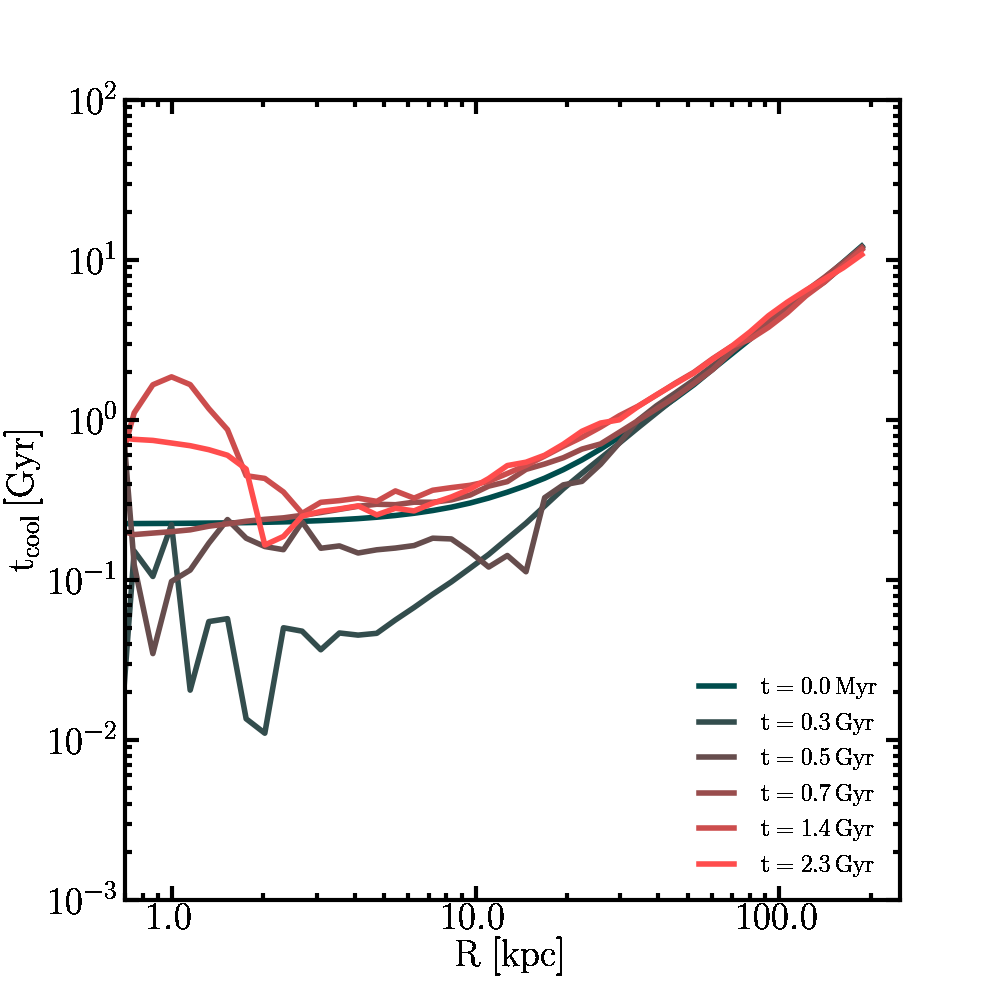}
\end{center}
\caption{ Cooling time profiles of the gas with temperature ${\rm T>10^6 \, K}$ at different times in the ${\rm HR\_JET\_HLLC}$ simulation which adopts a purely kinetic jet and has spatial resolution $\sim 200$ pc. }\label{fig:tcool}
\end{figure}

{The multi-phase structure of the gas can be appreciated in the density/temperature maps of Figure~\ref{fig:maps} and \ref{fig:mapsT}, as well as in the fluctuations of the mass-weighted gas density/temperature profiles of Figure~\ref{fig:profilesm}. In practice, dense gas almost always sits near the temperature floor at temperature ${\rm T< 10^6 \, K}$. This `cool' gas occupies a very small fraction of the volume and does not participate in the global cooling flow, because it cannot cool below the temperature floor. For this reason, to more meaningfully summarize the properties of the global cooling flow, we plot the cooling time profile of gas with temperature ${\rm T>10^6 \, K}$ in Figure~\ref{fig:tcool}. This figure shows how after the initial development of a strong cooling flow with short central cooling times, the ICM settles to a quasi-steady cooling time profile similar to that in the initial conditions. }

\begin{figure*}
\begin{center}
    \includegraphics[width=0.9\textwidth]{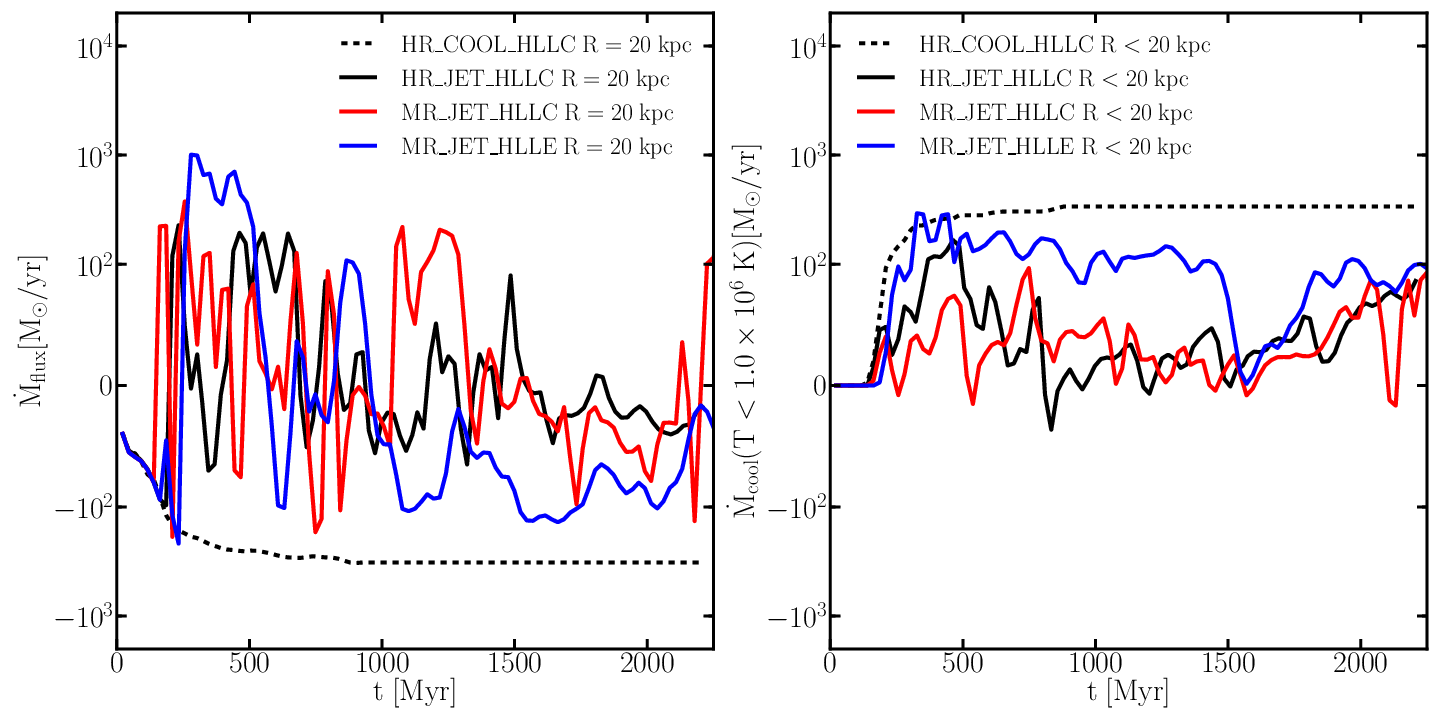}
\end{center}
\caption{ Comparison of runs with different jet physics, resolution and hydro solver. Left: total inward mass flux; negative values indicate inflow, positive values indicate outflow. Right: deposition rate of cool gas; positive values indicate cooling of gas from the `hot' to the `cool' phase, negative values indicate heating of gas from the `cool' to the `hot' phase. Mass flux and deposition rate are both reduced when jet feedback is turned on. The mass deposition rate and the inflow rate are very similar for runs with HLLC at different resolution.}\label{fig:resolution_mflux}
\end{figure*}

{To better understand the result of our simulations, we analyse how gas is transported and how it cools.} The left panel of Figure~\ref{fig:resolution_mflux} shows the radial mass flux at $R=20$ kpc. Positive values indicate a net outflow, negative values indicate a net inflow. The figure demonstrates that simulations with a fully kinetic jet are able to reduce the net mass inflow of gas towards the central regions relative to cool core clusters without jet heating. The same conclusion is reached by both our low resolution and high resolution runs. At fixed resolution, quantitative differences are noticeable comparing the results of the HLLE and HLLC solvers, but the qualitative picture does not change. The right panel of Figure~\ref{fig:resolution_mflux} shows the deposition rate of gas from the `hot' phase ($T\geq 10^6$ K) to the `cool' phase ($T<10^6$ K) within $R<20$ kpc. In principle, the amount of `cool' gas within this region can increase because of (I) cooling and (II) transport of `cool' gas from outside. Only the contribution from cooling is considered in this plot. Positive values are associated with increases in the gas cooling rate from the `hot' to the `cool' phase. Negative values are associated with heating from the `cool' phase to the `hot' phase. It is evident that simulations with a fully kinetic jet suppress the net cooling rate by a factor $\sim 10-100$. It is encouraging to see that for simulations with HLLC the results do not vary much when the resolution is increased. Cooling suppression is achieved also with HLLE, but the effect is somewhat weaker. The difference in the results may be explained by the inability of HLLE to maintain contact discontinuities between `hot' and `cool' gas, which is better achieved by HLLC and which influences the survival rate of cold clumps/filaments where the cooling rate is high. 

\begin{figure*}
\begin{center}
    \includegraphics[width=0.49\textwidth]{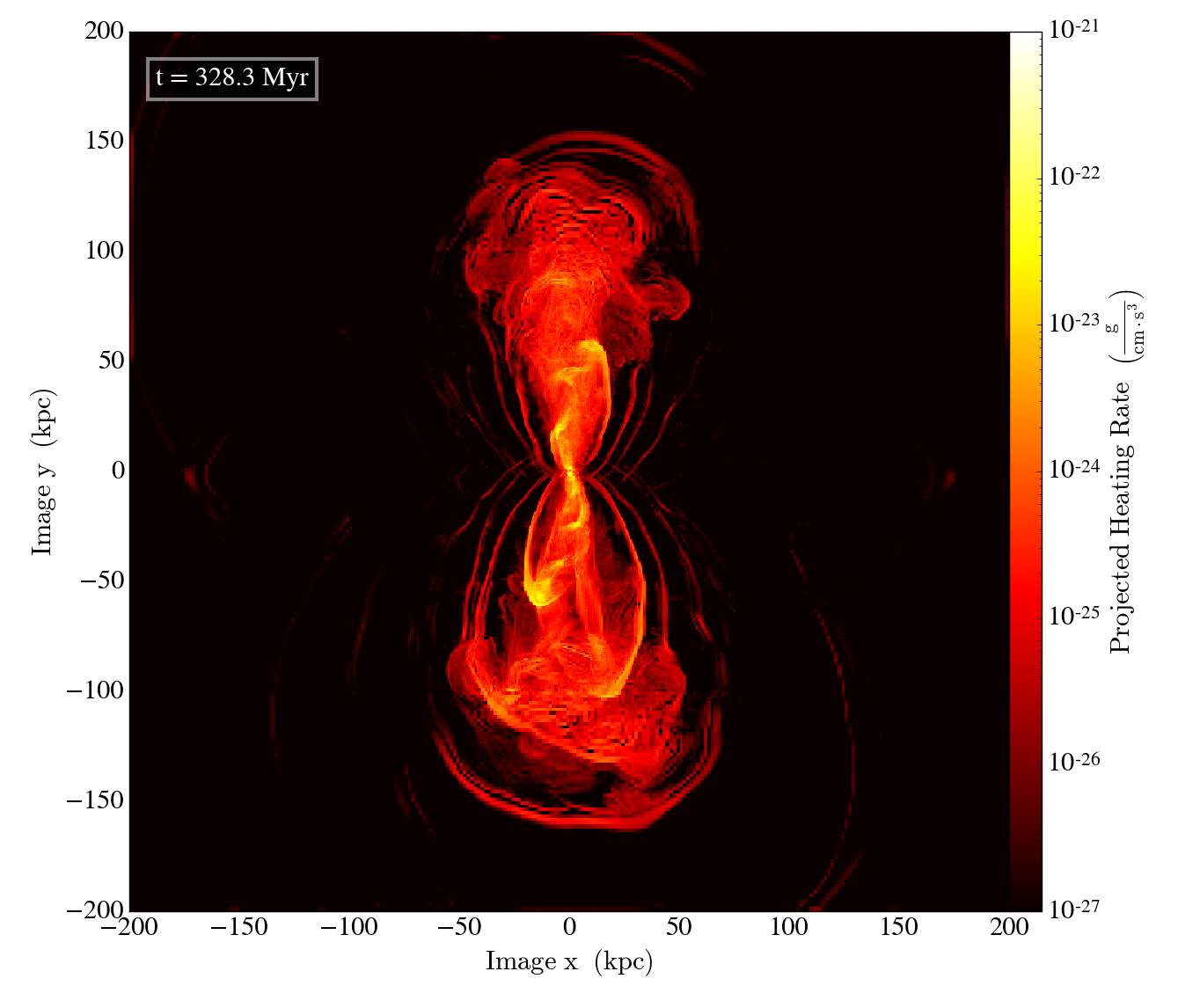}
    \includegraphics[width=0.49\textwidth]{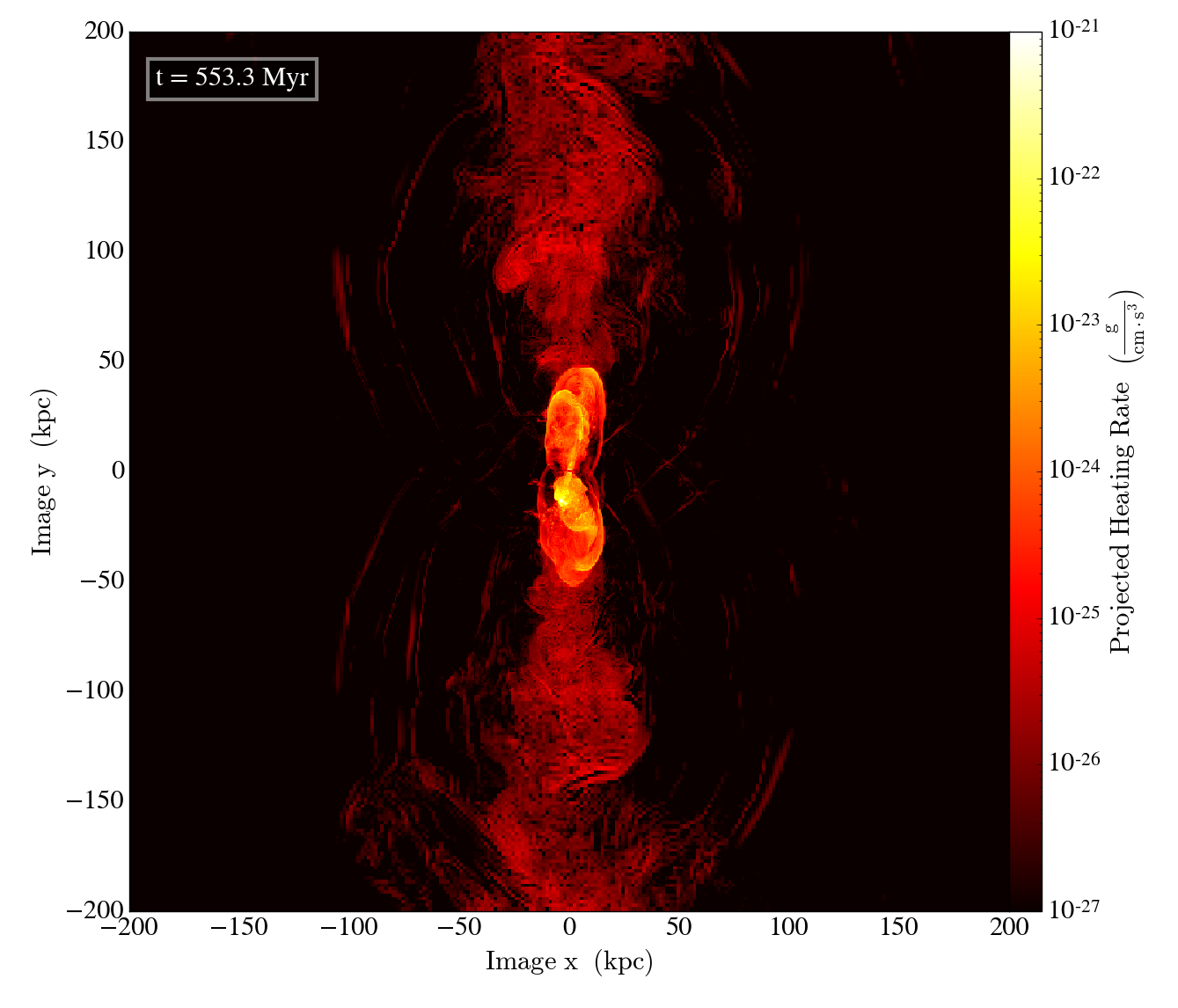}
    \includegraphics[width=0.49\textwidth]{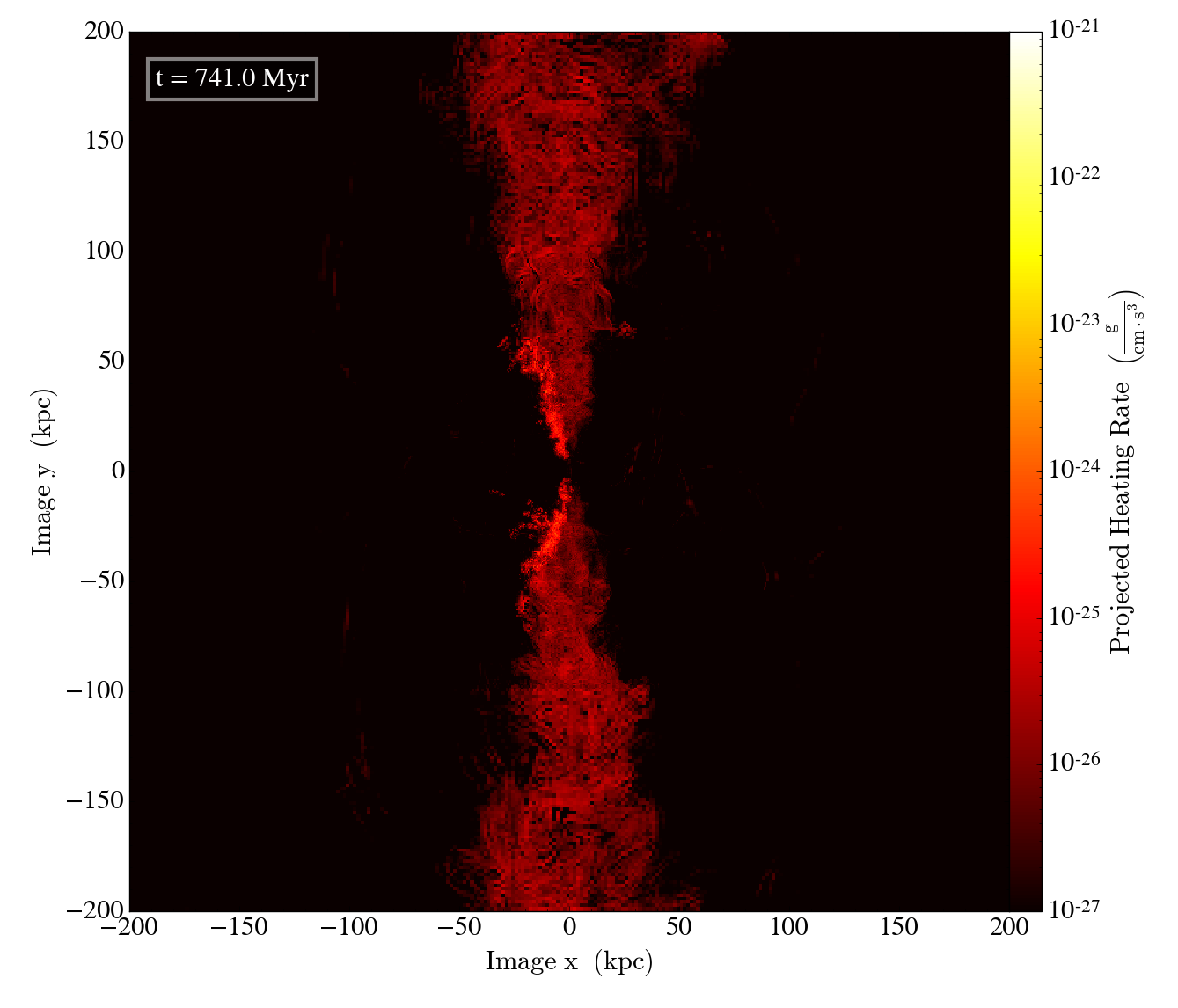}
    \includegraphics[width=0.49\textwidth]{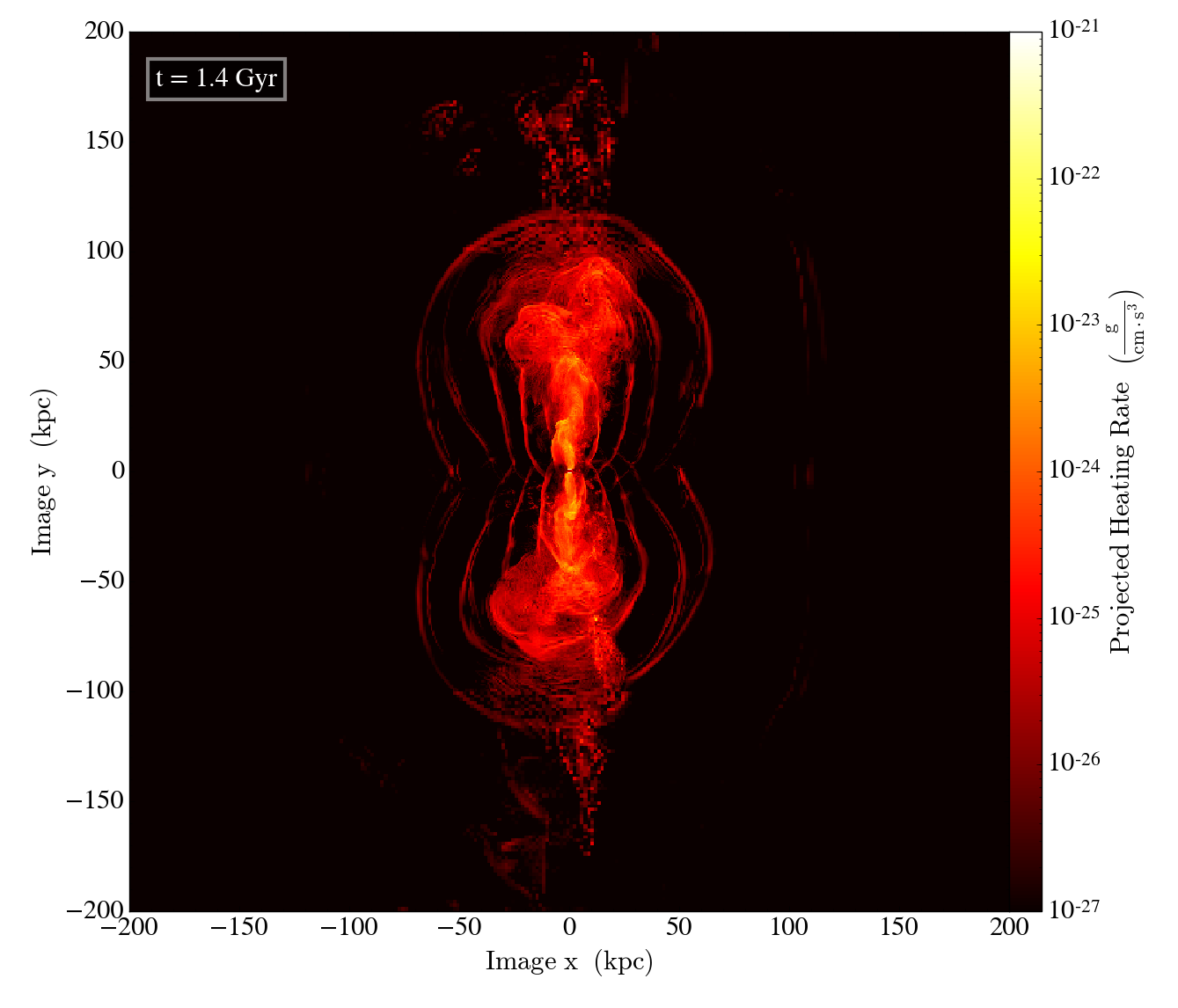}
\end{center}
\caption{ Average heating rate of the gas in thin slices at four different times for the ${\rm HR\_JET\_HLLC}$ simulation. The thickness of each slice is 5 kpc. We have chosen the same snapshots as in Figure 1 for clarity. Heating is very anisotropic and happens mostly along the jet axis. Heating away from the jet axis is achieved through weak shocks.}\label{fig:maps_heat}
\end{figure*}

The heating diagnostic described in Subsection~\ref{sec:tot_heat} is used to plot the average heating rate in thin slices in Figure~\ref{fig:maps_heat}. We show exactly the same snapshots as in Figure~\ref{fig:maps}. At all times heating is very anisotropic and it is distributed mostly along the jet axis and within the cavities. When the jet is fully on (top-left, bottom-right panels) heating appears to be distributed in shells which propagate outwards. These shells are generated when the jet sweeps through a region, then points towards another direction due to its precession. The heating shells propagate outwards as weak shocks, which constitute the main heating source away from the jet axis (see text below). 

\begin{figure*}
\begin{center}
    \includegraphics[width=0.9\textwidth]{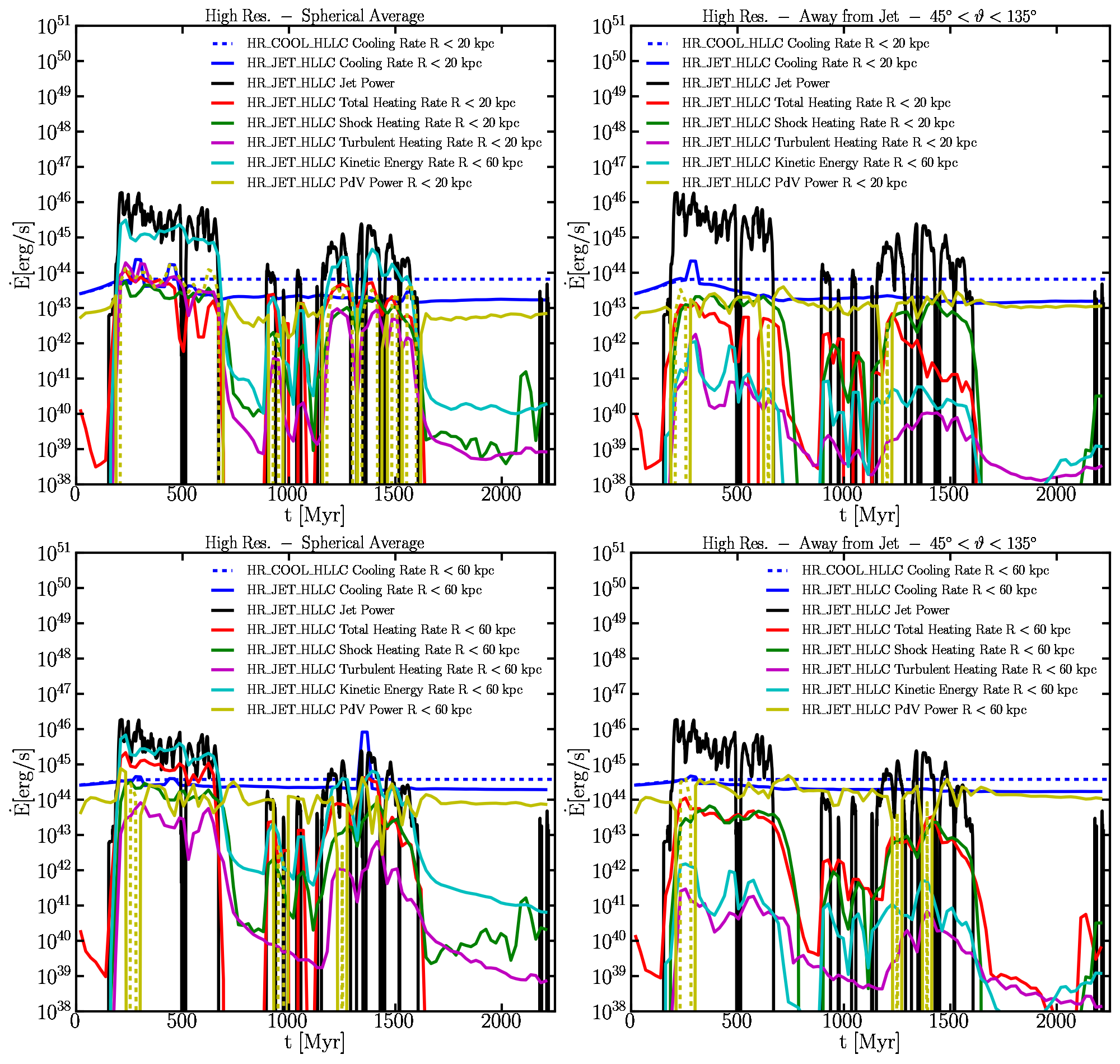}
\end{center}
\caption{ Temporal evolution of the cooling and heating rates within a spherical region of radius $R < 20$~kpc (top panels) and $R < 60$~kpc (bottom panels) for simulations with purely kinetic jets. The blue dashed line is the cooling rate in the cooling-only runs.  The blue solid line is the cooling rate in the runs with jet turned on. {The cooling rate only includes the contribution from gas at temperature ${\rm T\geq 10^6 \, K}$, i.e. it is the cooling rate of the gas cooling from the `hot' phase to the `cool' phase.} The black solid line is the jet power. The red solid line is the total heating rate. The green solid line is an estimate of the shock heating rate which is most accurate away from the jets, but is only a lower limit in the jet cone. {The cyan line represents the power associated with radial gas motions.} The magenta line is an estimate of the turbulent heating rate. {The solid yellow line represents the $PdV$ power when it is positive, whereas the dashed yellow line shows the (absolute value of the) $PdV$ power when it is negative.} {Most of the jet is used to accelerate gas into radial outflow motions within the jet cone.} Jet heating is very efficient at $R < 20$~kpc and the cooling rate is significantly decreased with respect to the value measured in the simulation without a jet. The total power injected by the jet is distributed out to $R \sim 60$~kpc, but the heating rate away from the jet axis is small. At large radii and away from the jet most heating is provided by weak shocks. The turbulent heating rate is significant only at $R < 20$~kpc, but it is always sub-dominant with respect to shock heating at larger radii. {$PdV$ power typically becomes significant only when the jet turns off and is then comparable to the cooling rate at all radii examined, a signature of a cooling flow (Appendix~\ref{appendix:C}).}} \label{fig:fiducial_heatcool}
\end{figure*}

{Figure~\ref{fig:fiducial_heatcool} shows the evolution of cooling/heating rates in regions of radius $R < 20$~kpc and $R < 60$~kpc, respectively. The cooling rate shown in this figure only includes the contribution from gas at temperature ${\rm T\geq 10^6 \, K}$, i.e. it is the cooling rate of the gas cooling from the `hot' phase to the `cool' phase. We consider both the spherically integrated case and the values from a region excluding the jet material ($45^\circ < \vartheta < 135^\circ$). The cooling rate within $R < 20$~kpc (top panels of Figure~\ref{fig:fiducial_heatcool}, solid blue lines) is smaller by a factor $\sim 10$ with respect to the case without jet heating (dashed blue lines), and a decrease of a factor $\sim 5$ is also achieved over bigger regions ($R < 60$~kpc, bottom panels of Figure~\ref{fig:fiducial_heatcool}, blue lines). The red solid lines in Figure~\ref{fig:fiducial_heatcool} show the total heating rate measured using the total heating diagnostic (Subsection~\ref{sec:tot_heat}), which can be compared to the jet power (black solid line) and to the kinetic power associated with radial outflows (cyan solid line). The yellow lines in Figure~\ref{fig:fiducial_heatcool} show the $PdV$ power. Within radii $R < 20$~kpc only  1-10\% of the total energy provided by the jet is converted to heat (top left panel), but when we consider the larger region $R < 60$~kpc (bottom left panel) the heating rate and the jet power match more closely, a signature of the fact that a larger fraction of the jet power has been converted into heat on scales $\gtrsim 50$ kpc. In the spherically  averaged sense (left panels), the heating rate closely balances the cooling rate when the jet is active at maximum power. Nonetheless, it appears that most of the power provided by the jet is used to accelerate gas into radial outflows within the jet cone. This phenomenon prevents gas from flowing back to the cluster centre, with the effect of maintaining central cooling rate suppression even if the jet turns off. The combination of these effects causes a reduction of the cooling rate in the central regions ($R<20$~kpc), whereas the system relaxes to a cooling flow configuration at larger scales ($R<60$~kpc).}

Heating is very anisotropic: if a cone of amplitude $45^\circ$ containing the jet is excluded from the analysis (right panels of Figure~\ref{fig:fiducial_heatcool}), we see that the total heating rate is much smaller than the cooling rate and than the jet power. In the central region ($R < 20$~kpc, top-left panel of Figure~\ref{fig:fiducial_heatcool}) the turbulent heating rate (magenta line) is comparable to the total heating rate (red line) when the jet is on. On larger scales ($R < 60$~kpc, bottom-left panel of Figure~\ref{fig:fiducial_heatcool}) the turbulent heating rate only constitutes $\sim 1\%$ of the total heating rate. Green lines in Figure~\ref{fig:fiducial_heatcool} show the shock heating rate which is sub-dominant at $R < 20$~kpc, but dominates the heating rate at larger scales ($R < 60$~kpc). In particular, when the jet region is excluded from the analysis (right panels of Figure~\ref{fig:fiducial_heatcool}) shock heating represents the largest contribution to the total heating rate. 

{From Figure~\ref{fig:fiducial_heatcool} it appears that $PdV$ power is typically significant when the jet turns off and the ICM starts responding adiabatically to the perturbations generated by the jet. Away from the jet, the $PdV$ power is comparable to the cooling rate both at radius $R < 20$~kpc and $R < 60$~kpc, which is a signature of a cooling flow (see Appendix~\ref{appendix:C})}

{Figure 4-9 demonstrate (I) that our fiducial simulations with a purely kinetic jet, with the HLLC Riemann solver are successfully reducing the central cooling rate and reduce the central cooling flow of a Perseus-like cool-core cluster, (II) that $50-100$\% of the heating on small scales is provided by turbulent energy dissipation, but that contribution is $\sim1\%$ at larger scale, (III) that shock heating is the dominant heating source on large scales and away from the jet axis, (IV) that most of the jet power is used to accelerate gas into a conical radial flow which prevents gas from falling back to the cluster centre when the jet is off, and (V) that $PdV$ power is comparable to the cooling rate which is evidence for a reduced cooling flow. The fact that the system relaxes to a reduced cooling flow is in broad agreement with the qualitative picture that has emerged from recent work published using similar hydrodynamical simulations \citep{2014ApJ...789...54L, 2015ApJ...811...73L, 2016ApJ...829...90Y, 2017ApJ...847..106L, 2017ApJ...841..133M}.} The next few subsections discuss our results from non-fiducial runs which highlight several differences with respect to the existing literature that we largely attribute to our adoption of less diffusive Riemann solvers. 

\subsection{Purely Kinetic Jet vs. Mixed Injection}\label{sec:kin_therm}

\begin{figure*}
\begin{center}
    \includegraphics[width=0.9\textwidth]{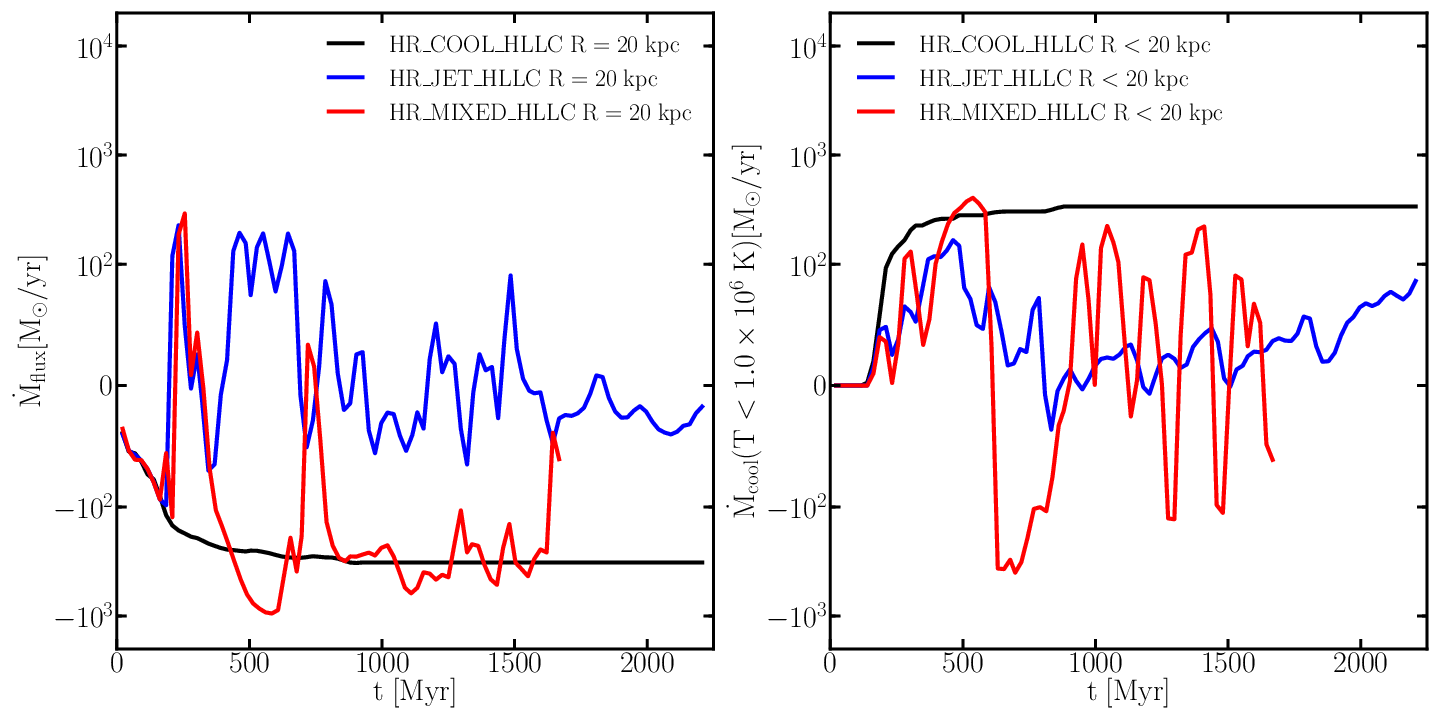}
\end{center}
\caption{ Comparison of purely kinetic vs. mixed feedback scheme in the high resolution runs. Left: total inward mass flux at 20 kpc; negative values indicate inflow, positive values indicate outflow. Right: deposition rate of cold gas within 20 kpc; positive values indicate cooling of gas from the ``hot'' to the ``cold'' phase, negative values indicate heating of gas from the ``cold'' to the ``hot'' phase. Only simulations with purely kinetic jets can simultaneously regulate the mass inflow rate and reduce the deposition rate from the ``hot'' to the ``cold'' phase.}\label{fig:kin_vs_mix_mflux}
\end{figure*}

Figure~\ref{fig:kin_vs_mix_mflux} shows the mass flux at $R=20$ kpc (left panel) and the deposition rate of gas from the `hot' phase ($T\geq 10^6$ K) to the `cool' phase ($T<10^6$ K) within $R<20$ kpc (right panel) for our high resolution simulations with cooling only, purely kinetic jet and mixed thermal+kinetic injection. This figure shows that our case with mixed injection struggles to suppress the total flow of gas towards the centre of the cluster, and that there are larger fluctuations of the `cool' gas deposition rates with respect to the case with purely kinetic jet. 

\begin{figure*}
\begin{center}
    \includegraphics[width=0.9\textwidth]{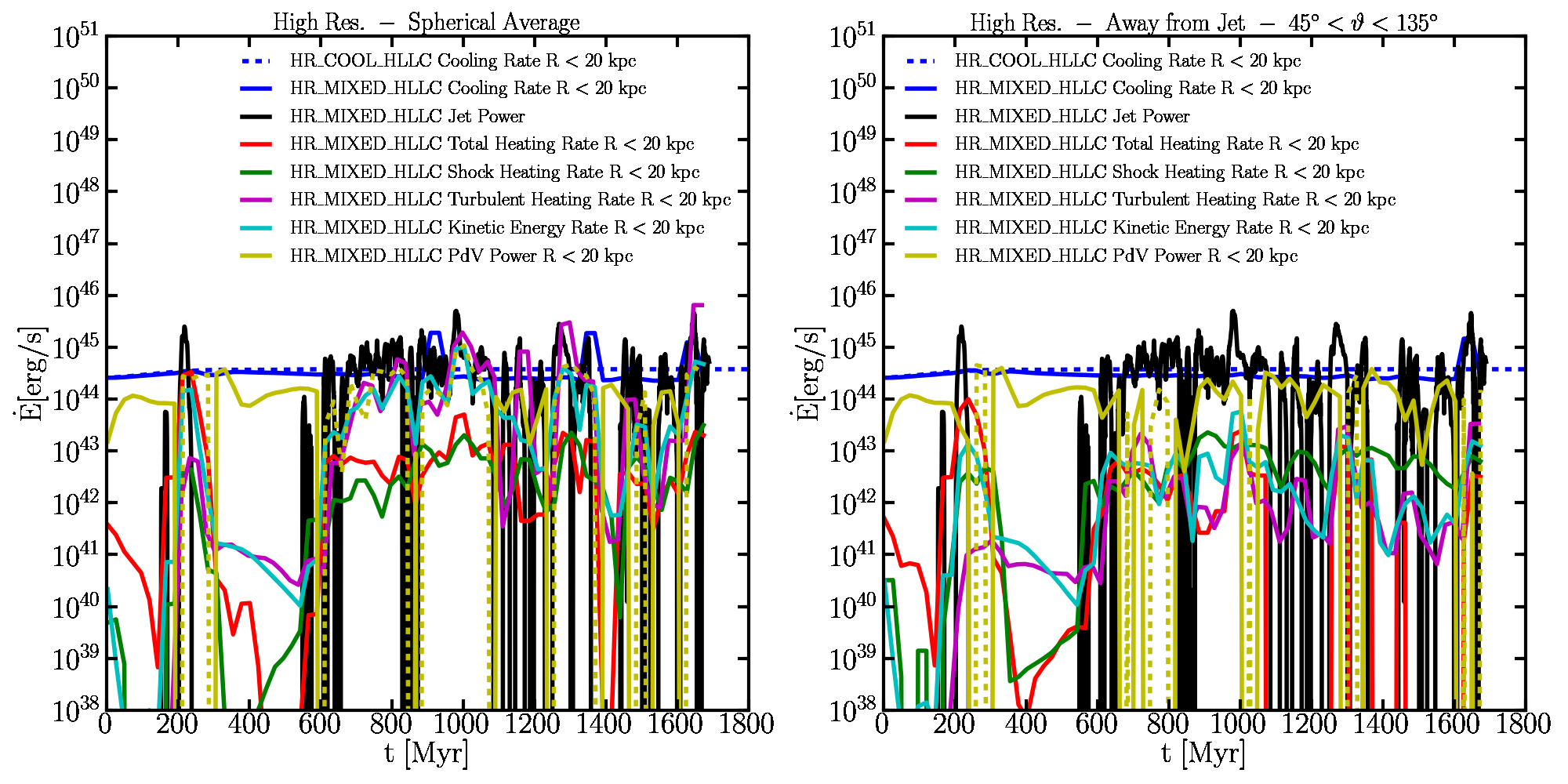}
\end{center}
\caption{ Temporal evolution of the cooling and heating rates within a spherical region of radius $R < 20$~kpc for simulations with mixed kinetic and thermal feedback. The blue dashed line is the cooling rate in the cooling-only runs. The blue solid line is the cooling rate in the runs with jet turned on. {The cooling rate only includes the contribution from gas at temperature ${\rm T\geq 10^6 \, K}$, i.e. it is the cooling rate of the gas cooling from the `hot' phase to the `cool' phase.} The black solid line is the jet power. The red solid line is the total heating rate. The green solid line is an estimate of the shock heating rate which is most accurate away from the jets, but is only a lower limit in the jet cone. {The cyan line represents the power associated with radial gas motions.} The magenta line is an estimate of the turbulent heating rate. {The solid yellow line represents the $PdV$ power when it is positive, whereas the dashed yellow line shows the (absolute value of the) $PdV$ power when it is negative.} This plot shows that the mixed feedback is unable to regulate the cooling flow.}\label{fig:mix_heatcool}
\end{figure*}

Figure~\ref{fig:mix_heatcool} shows the evolution of heating and cooling rates at $R < 20$~kpc for simulations with mixed kinetic and thermal feedback. Comparison to Figure~\ref{fig:fiducial_heatcool} (purely kinetic jet) shows that our simulations with mixed injection fails at suppressing the cooling flow and reducing the central cooling rate. 

We conclude that the qualitative behavior of our simulations depends on the fraction of kinetic/thermal energy injected by the jet. In particular, injection of thermal energy instead of kinetic energy seems to significantly reduce the efficiency of AGN feedback. This is a well known result from cosmological zoom-in simulations that reach $\sim 0.5-1$~kpc where AGN feedback is performed by injection of pure thermal energy or a mixture of thermal and kinetic energy \citep{2009MNRAS.398...53B, 2012MNRAS.420.2662D, 2014MNRAS.441.1270L, 2017MNRAS.470..166H}. A direct comparison of our results to cosmological simulations is difficult to make, because of the very different way energy/momentum injection are implemented. For instance, the AMR simulations of \cite{2012MNRAS.420.2662D} inject kinetic energy jets in cylinders, but thermal energy in spheres, which is different from the approach followed to model AGN feedback in idealised simulations. In general, implementations of AGN feedback in cosmological simulations require thermal energy not to be released immediately, but rather accumulated and released in powerful, impulsive events. 

The theoretical work by \cite{2017ApJ...845...80V} has shown that it is possible to explain the multi-phase structure of gas in cluster cores while simultaneously achieving thermal balance. In this scenario, AGN feedback uplifts gas to large scales and is regulated by the condensation and precipitation of cold material formed via thermal instability. Typical power law entropy profiles with a flat central core are a natural outcome of this class of models. Condensation of cold gas is promoted in regions where the entropy profile is flat. At larger radii, where the entropy profile is steeper, condensation of cold gas is suppressed by buoyancy. \cite{2017ApJ...845...80V} argue that numerical methods based on pure thermal feedback fail to reach a precipitation-regulated regime, because thermal feedback generates an inversion of the slope of the central density profile which promotes the formation of large quantities of cold gas in the central regions. In these conditions, the core is unable to reach thermal balance. The phenomenon may be alleviated if some of this material can be lifted up and transported towards buoyantly unstable regions away from the centre via kinetic feedback. For this reason, \cite{2017ApJ...845...80V} argue that injecting at least a fraction of AGN feedback energy into a kinetic jet is important to achieve regulation. These arguments may partially explain why our simulations with purely kinetic jet regulates the cooling flow better than those with mixed feedback, but they do not explain why our mixed feedback simulations fail so dramatically. 

Our results are in contrast with numerical work by other authors who also varied the fraction of thermal energy injected by the jet in idealised simulations \citep{2014ApJ...789...54L, 2017ApJ...841..133M}. \cite{2017ApJ...841..133M} ran extensive tests with resolution comparable to ours and concluded that their simulations achieve the same qualitative behavior when the injected kinetic energy fraction is sufficiently larger than zero. Adoption of coarse resolution might significantly influence thermalization of kinetic energy and might cause overcooling, which is artificially prevented by energy accumulation in cosmological zoom-in simulations. However, the resolution in our simulations and in \cite{2017ApJ...841..133M} is not significantly different from the highest resolution achieved by cosmological zoom-in simulations ($\sim 0.5$ kpc), so the discrepancy between \cite{2017ApJ...841..133M} and results form cosmological simulations (and ours) cannot only be attributed to resolution effects. 

\cite{2014ApJ...789...54L} performed simulations at very high resolution and showed that the results do not depend significantly on the injected kinetic energy fraction provided that the resolution is at least a few hundred parsec, in tension with our results. In principle, it should be possible to assess whether the kinetic/thermal fraction matters by significantly increasing the spatial resolution, hoping that the correct physics would be explicitly captured by the hydro method. However, this would only be true if all numerical methods converged at the same rate towards the appropriate solution as the resolution is increased, which is not the case even for the high resolution runs performed nowadays. 

The simulations by \cite{2014ApJ...789...54L} and \cite{2017ApJ...841..133M} have one aspect in common, which is also the major difference with respect to our setup: they use the {\sc zeus} solver \citep{1992ApJS...80..753S} implemented in the {\sc enzo} code \citep{2014ApJS..211...19B}. The {\sc zeus} method introduces significantly larger numerical diffusion compared to methods such as HLLE or HLLC, which also allowed these authors to have better numerical stability and run simulations with lower temperature floor than we achieve. In particular, \cite{2017ApJ...841..133M} state that ``{\sc zeus} is known to be a relatively diffusive method and requires an artificial viscosity term that may affect the accuracy of our hydrodynamics calculations.'' Furthermore, \cite{2017ApJ...841..133M} have experimented with using a piecewise parabolic method (PPM), but encountered numerical difficulties relating to the strong discontinuities occurring at the injection site. The extra numerical diffusion of {\sc zeus} may (I) achieve faster thermalization of kinetic energy and (II) smooth out large density and temperature gradients at the interface of clumps/filaments formed via thermal instability that play an important role in the duty cycle of AGN activity. Our simulations with HLLC are expected to capture discontinuities better than simulations with {\sc zeus}, and are less affected by spurious thermalization of kinetic energy.  

\cite{2016ApJ...829...90Y} used the {\sc flash} code \citep{2000ApJS..131..273F} and adopted the directionally unsplit staggered mesh solver (USM, \citealt{2009ASPC..406..243L, 2013JCoPh.243..269L}). The numerical scheme used by \cite{2016ApJ...829...90Y} is similar to ours, but their minimum cell size is a factor $\sim 10$ larger than in our HR case. In contrast to our simulations, \cite{2016ApJ...829...90Y} used a sub-resolution method that replaces `cold' gas below a temperature threshold with particles, which are also used to compute the accretion rate. Finally, \cite{2016ApJ...829...90Y} do not directly address the issue of thermal vs. kinetic energy injection. For these reasons a direct comparison to their work is difficult to make.

In general, it is possible that issues other than numerical diffusion may influence the results. For instance, in models that use cold gas accretion the jet power can be larger using a larger accretion region, which will produce stronger heating of the cluster core. Additionally, the choice for the temperature floor influences the density structure of the `cool' gas formed in the cluster core which determines the black hole feeding rate and the jet power. Because of the dependence of the density distribution of the `cool' gas on the temperature floor, the coupling efficiency of kinetic jets and thermal energy with this gas phase can be influenced.

\begin{figure*}
\begin{center}
    \includegraphics[width=0.9\textwidth]{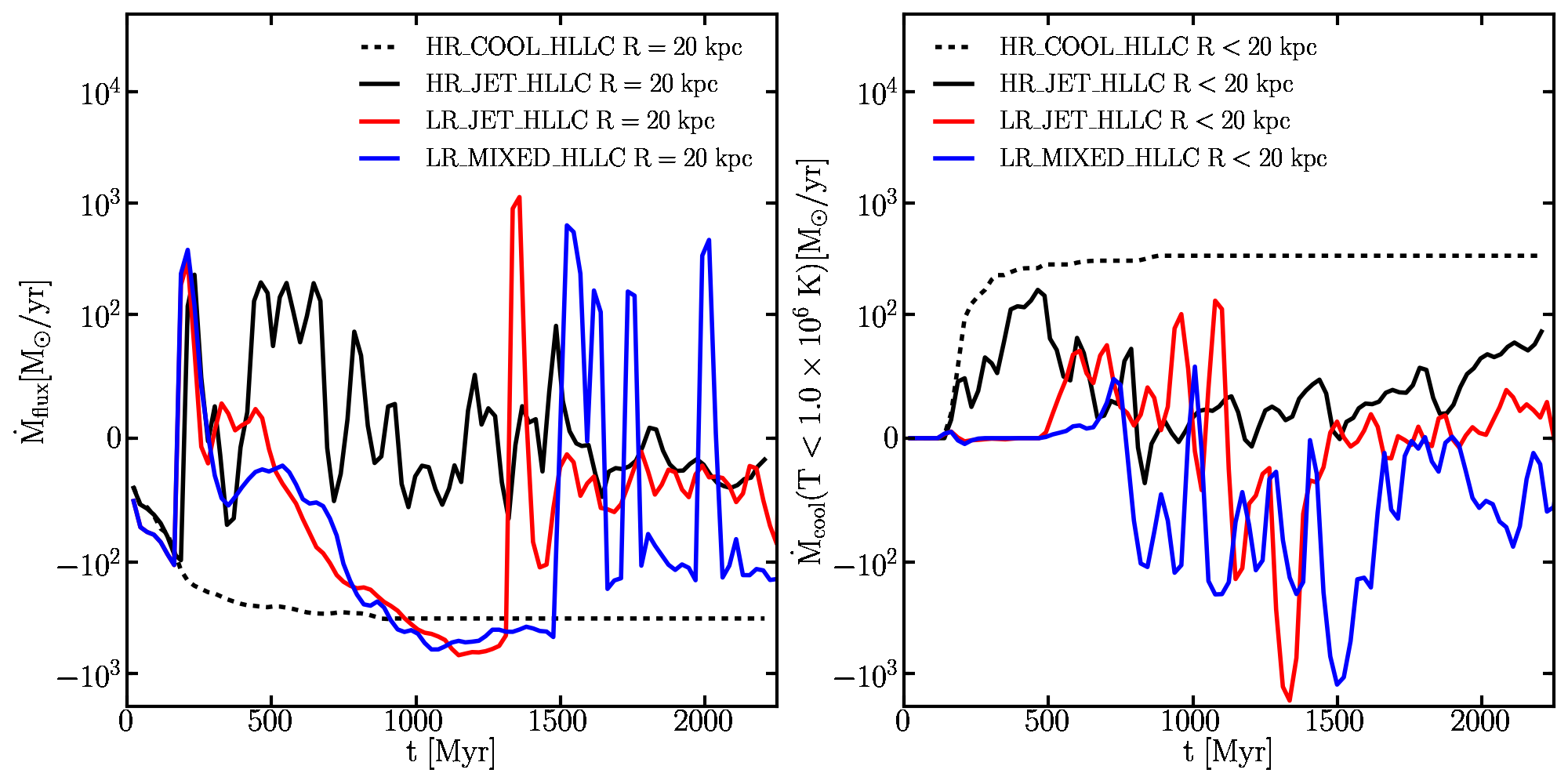}
\end{center}
\caption{Comparison of purely kinetic vs. mixed feedback scheme in the low resolution (LR) runs. Left: total inward mass flux at 20 kpc; negative values indicate inflow, positive values indicate outflow. Right: deposition rate of cold gas within 20 kpc; positive values indicate cooling of gas from the ``hot'' to the ``cold'' phase, negative values indicate heating of gas from the ``cold'' to the ``hot'' phase. Results from the LR runs are compared to the fiducial HR run with purely kinetic jet. At low resolution the purely kinetic jet simulation and the mixed feedback simulation converge to similar solutions, unlike what happens at higher resolution. }\label{fig:kin_vs_mix_mflux_lr}
\end{figure*}

Nevertheless, our result that numerical diffusion affects the dependence of jet feedback on the kinetic/thermal energy fraction is strengthened by an additional test. We ran two simulation with the HLLC solver at low resolution LR ($\sim 1.5$ kpc) with purely kinetic jet and mixed thermal/kinetic feedback, respectively. Figure~\ref{fig:kin_vs_mix_mflux_lr} shows the comparison of the total inward mass flux at 20 kpc and the deposition rate of cold gas within 20 kpc for these two low resolution runs. We find that at resolution $\gtrsim 1.5$ kpc, simulations with purely kinetic jet and mixed feedback converge to similar solutions, unlike what happens at higher resolution. 

In summary, it appears that the dependence of the results on the kinetic/thermal energy fraction in the jet is sensitive to the details of the numerical setup, including the diffusivity of the hydro solver and the specific criteria used for gas accretion and the method of jet injection. Given the difficulties in obtaining numerically converged solutions for this problem using accurate hydro solvers, existing results on heating of the ICM by jets (including ours) should be interpreted with the proper caveats and regarded as provisional. 

\subsection{Different Riemann Solvers and Robustness of Numerical Solutions}\label{sec:rsolvers}

\begin{figure}
\begin{center}
    \includegraphics[width=0.45\textwidth]{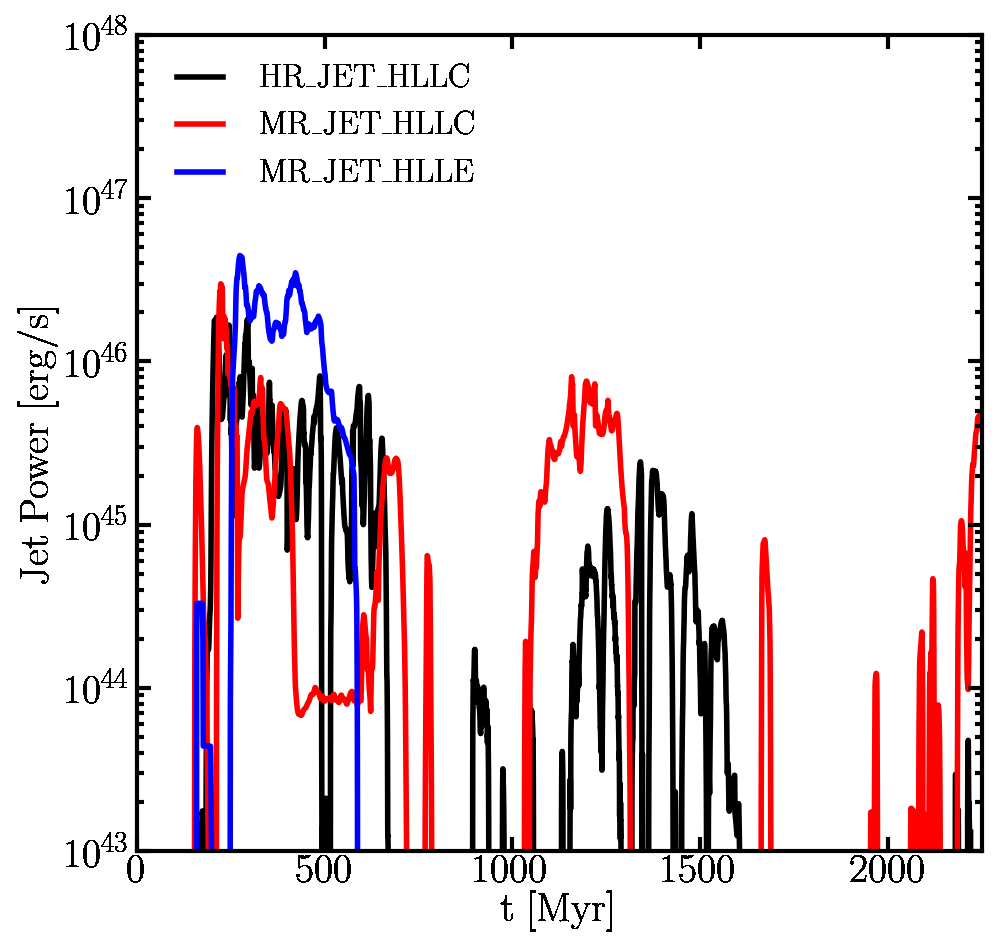}
\end{center}
\caption{ Jet power in simulations with different jet physics, resolution and hydro solver. The supply of gas that triggers the jet is sensitive to the choice of the Riemann solver. This result may reflect the chaotic nature of the system. }\label{fig:jet_power_solver}
\end{figure}

The physics that determines the gas supply that triggers jet events is determined by (I) the magnitude of the cooling flow, and (II) the properties and dynamics of the `mist' of cold clumps/filaments embedded in the hot ICM that are formed via thermal instability. The former can be easily captured by any state-of-the-art hydro code, but the latter cannot. In general, it is unclear what spatial resolution or what numerical solver is more suitable to capture a physically reliable picture of the thermal instability in galaxy cluster cores heated by a jet. Figure~\ref{fig:jet_power_solver} shows how the evolution of the jet power depends on numerical resolution and Riemann solver in a non-linear and unpredictable fashion. In particular, at fixed resolution and temperature floor, the HLLE and HLLC Riemann solvers predict very different evolutions of the jet power. By construction, HLLC better resolves multi-phase flows and captures contact discontinuities, but this test highlights how choosing a different solver can influence one aspect of the numerical solution. This result may reflect the chaotic nature of the system.

\begin{figure*}
\begin{center}
    \includegraphics[width=0.49\textwidth]{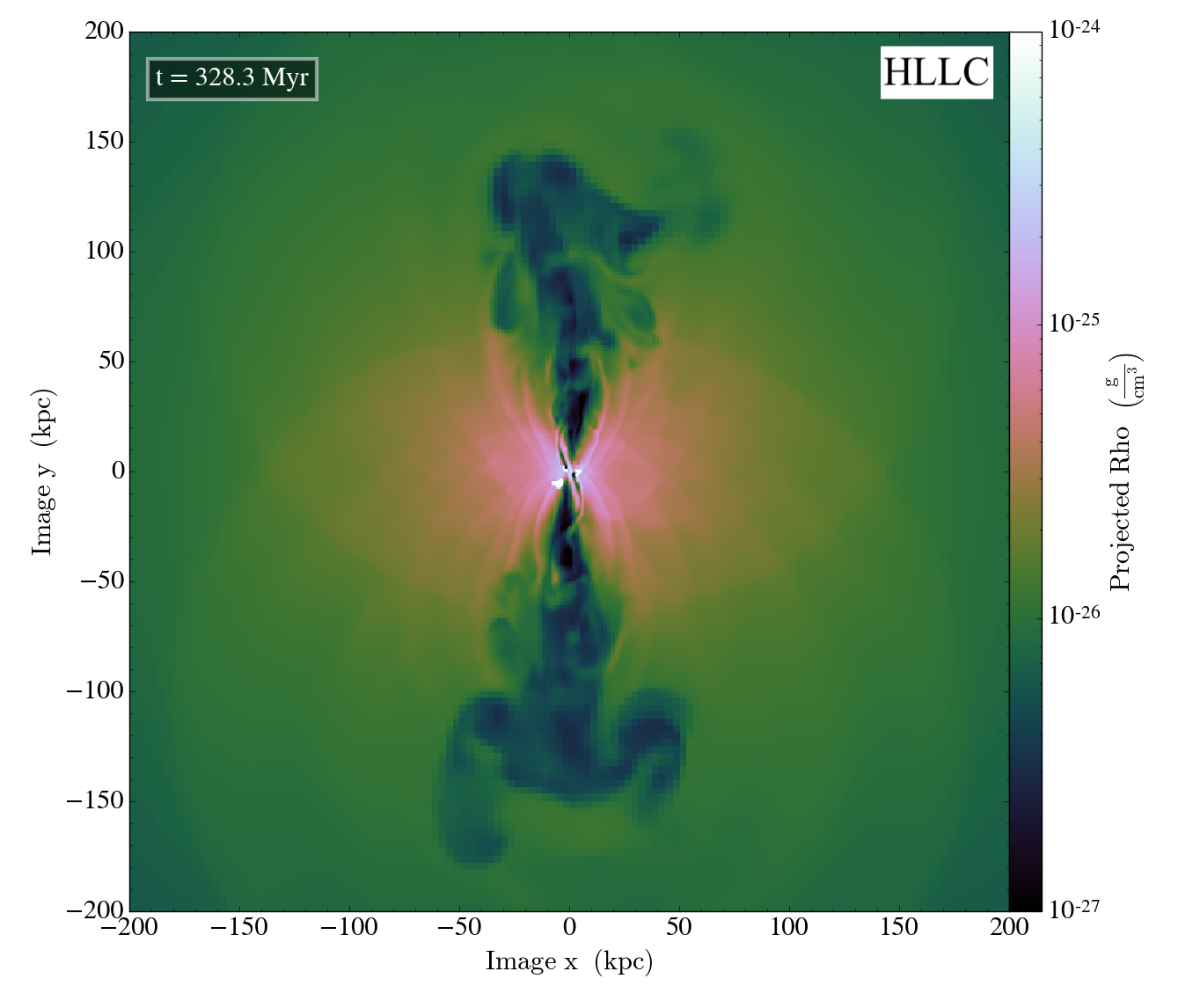}
    \includegraphics[width=0.49\textwidth]{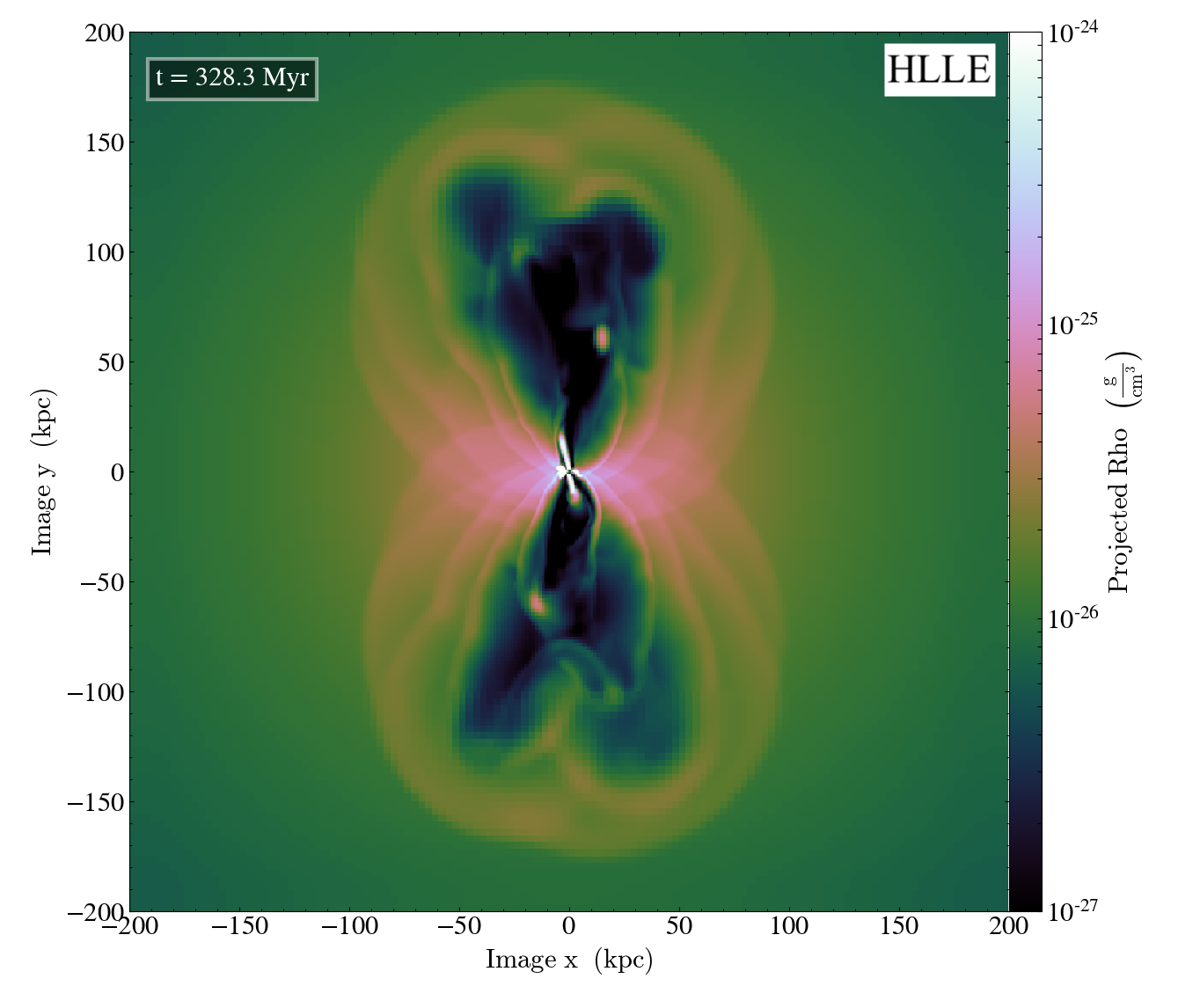}
    \includegraphics[width=0.49\textwidth]{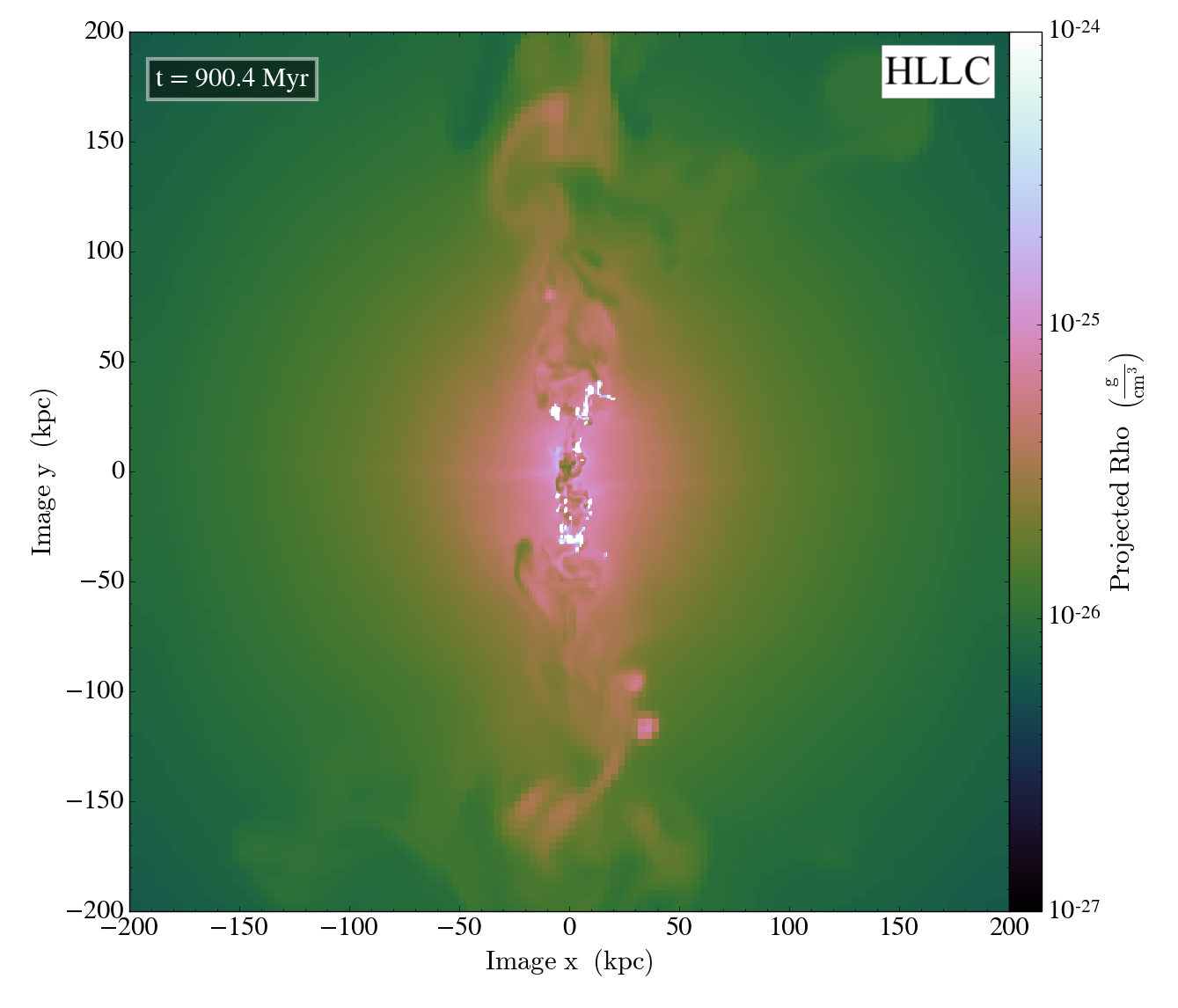}
    \includegraphics[width=0.49\textwidth]{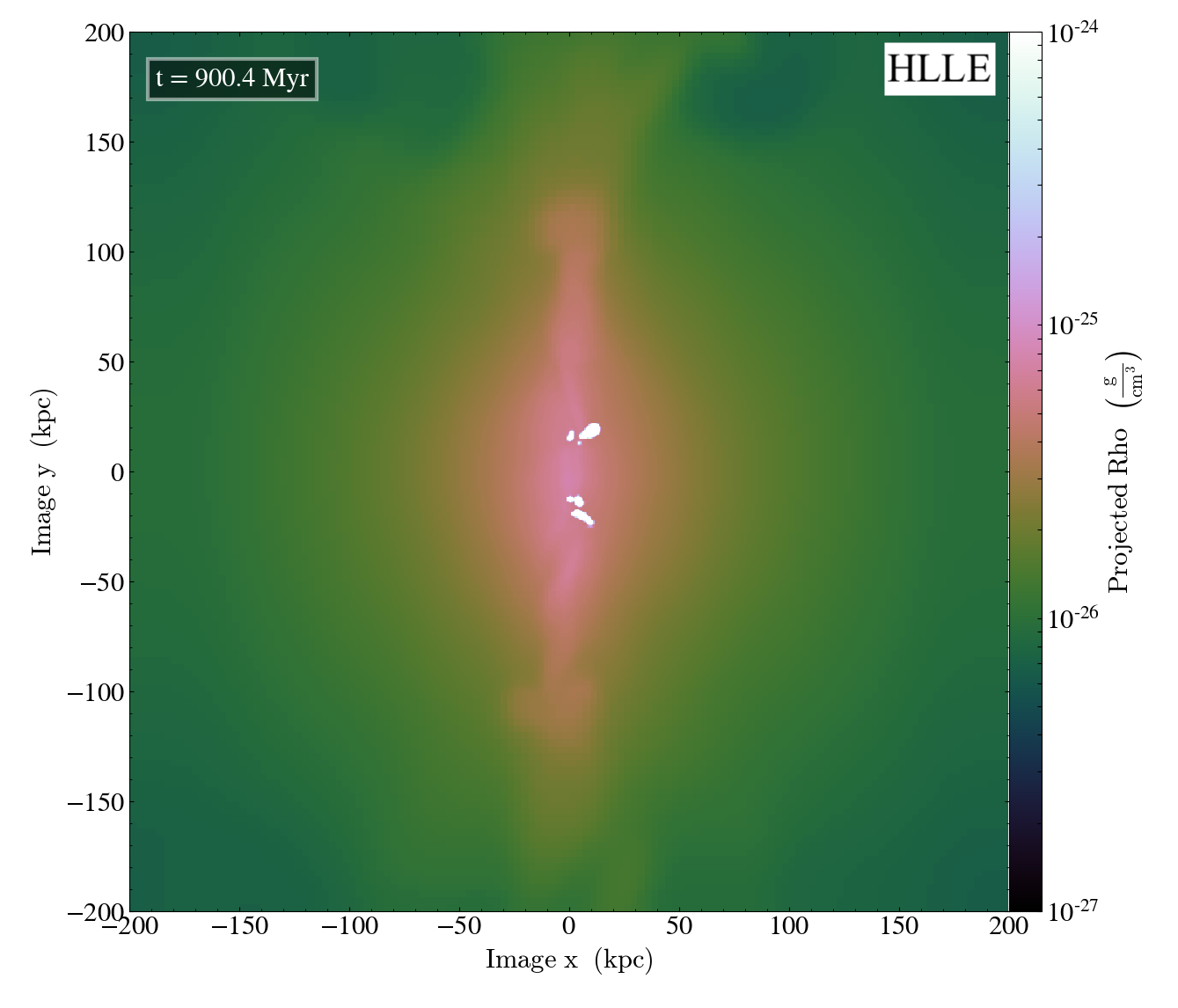}
\end{center}
\caption{ Average density of the gas in thin slices at four different times for the ${\rm MR\_JET\_HLLC}$ (left) and ${\rm MR\_JET\_HLLE}$ (right) simulations. The thickness of each slice is 5 kpc. See text for discussion. Top panels: snapshots chosen at times when the jet is on with similar power. The degree of gas mixing in the jet cavities is influenced by the chosen Riemann solver. Bottom panel: snapshots chosen at times when the jet has been off for more than 50 Myr. The formation of cold clumps and filaments is significantly influenced by the chosen Riemann solver. }\label{fig:maps_solver}
\end{figure*}

Figure~\ref{fig:maps_solver} shows a visual representation of the differences between HLLC and HLLE in similar phases of the ${\rm MR\_JET\_HLLC}$ and ${\rm MR\_JET\_HLLE}$ simulations. The top panels of Figure~\ref{fig:maps_solver} show density maps of the gas at times when the jet is on with similar power for HLLC and HLLE, respectively. It is evident that the degree of gas mixing in the cavities generated by the jet depends on the solver. In particular, HLLC appears to better preserve high density structures in the cavities. The bottom panels of Figure~\ref{fig:maps_solver} show density maps at times when the jet has been turned off for more than 50 Myr. In this phase, thermal instabilities develop which lead to the formation of cold clumps and filaments. The morphology of these structures appears quite different between HLLC and HLLE. In particular, HLLC allows the development of multiple high density structures, whereas the phenomenon is significantly less pronounced when HLLE is used.

These facts pose the question of whether a numerical solution that does not depend on the Riemann solver for the jet heating problem is achievable. In tests not shown here, we assess whether this strict kind of convergence can be achieved. In principle, it should be possible to obtain better numerical convergence by minimizing thermal instability at the resolution scale. In order to reach this goal, we artificially increase the temperature floor to values as high as ${\rm T_{floor}=10^7 \, K}$ and find that strict convergence between HLLC and HLLE is actually very hard to achieve. HLLE and HLLC converge to similar solutions only when we increase the temperature floor to ${\rm T_{floor}=10^7 \, K}$, but the similarities can only be appreciated if the results are time-averaged on time scales $\Delta t \sim 100$ Myr. The detailed evolution of the jet power depends on the Riemann solver and increasing the temperature floor only alleviates the differences between HLLE and HLLC. This may be a consequence of the chaotic nature of the system influencing the numerical solutions even if thermal instability is minimised. 

In conclusion, simulations with HLLC offer advantages when dealing with multi-phase structure, but the temperature floor cannot be decreased to desirable levels (${\rm T<5\times10^4 \, K}$) because of numerical instabilities. More diffusive solvers have the advantage of being more numerically stable when the temperature floor is lowered, but they introduce larger numerical errors. In principle, some of these issues could be alleviated by achieving very high resolution and fully resolve thermal instabilities on all relevant scales. Unfortunately, (I) the appropriate resolution requirement has not been clearly established from a theoretical viewpoint and it may be unachievable in state-of-the-art numerical simulations \citep{2018MNRAS.473.5407M}, and (II) the chaotic nature of the system may still produce different solutions for different numerical solvers, although the solutions should ideally converge statistically in time with sufficient time averaging.


\section{Summary and conclusions}

We have analysed the results of a suite of hydrodynamical simulations of AGN jet heating in galaxy cluster cores. The jets are treated in the non-relativistic limit. Our setup is similar to that adopted by other authors in recent years \citep{2012ApJ...746...94G, 2014ApJ...789...54L, 2016ApJ...829...90Y, 2017ApJ...847..106L, 2017ApJ...841..133M}, but our simulations adopt different numerical solvers. In particular, our fiducial simulations adopt the HLLC Riemann solver which offers lower numerical diffusion compared to solvers used previously in the literature and better handles multi-phase flows. The latter feature is important when dealing with thermal instabilities that generate clumps embedded in the hot ICM (contact discontinuities) and that play a key role in cooling flows heated by AGN feedback \citep{2012MNRAS.420.3174S}. 

We implemented jet feedback triggered by gas accretion onto a supermassive black hole at the centre of a cluster core and we considered both the scenario with a purely kinetic jet and the one with mixed kinetic and thermal energy injection. We also implemented an on-the-fly heating estimator inspired by \cite{2015MNRAS.454.1848R} which measures the total heating in each cell in the computational domain. The turbulent energy decay rate and the shock heating rate are also estimated using approximate methods. 

{ We found that our fiducial simulations with the HLLC Riemann solver and with purely kinetic jet injection are able to reduce the gas cooling rate in the central 20 kpc by a factor 10 with respect to cooling-only simulations. Most of the jet power is used to accelerate gas into conical radial outflows with only $\sim 10-50 \%$ going into heating. Jet heating is anisotropic and achieved mostly along the jet axis. Within the central 20 kpc, turbulent kinetic energy dissipation constitutes a significant fraction of the heating rate, but on larger scales its contribution is only a few percent of the total heating rate. Within 60 kpc a large fraction of the total heating rate is supplied via weak shocks generated by the jet. When the jet is on at full power, the cooling rate and the heating rate balance each other. When the jet turns off, the presence of the conical outflows prevents gas from rapidly flowing back to the central regions and establish a strong central cooling flow. As a result, the cooling rate within the central 20 kpc is reduced with respect to cooling-only simulations even when the jet temporarily switches off. At large radii $R\sim 60$~kpc and away from the jet cone, the effect of the jet is reduced and the system settles into a reduced cooling flow configuration. The formation of a reduced cooling flow in hydrodynamical jet simulations was also reported by \cite{2016ApJ...829...90Y}. In summary, the effect of the jet is sufficient to locally regulate the cooling flow (in the central 20 kpc), but insufficient to regulate it on a global scale.}

{\cite{2017ApJ...845...91H} suggested that the dominant source of heating in cluster cores is provided by mixing of hot bubbles inflated by the jet with the background ICM, which is not suggested by our results. This difference can be understood from the fact that the simulations of \cite{2017ApJ...845...91H} used jets with a large opening angle, which favors the inflation of quasi-spherical bubbles rather than the conical lobes observed in our simulations of jets with a small opening angle.} 

Contrary to previous work, we found that regulation of the cooling flow is not achieved when mixed thermal and kinetic jet injection is used. We attribute this discrepancy to differences in the numerical solvers used to run the simulations. For instance, \cite{2014ApJ...789...54L} and \cite{2017ApJ...841..133M} adopt solvers (or spatial resolutions) that generate larger numerical diffusion with respect to our setup with HLLC and spatial resolution of $\sim 200$~pc. Having larger numerical diffusion likely influences the way the jet kinetic energy is thermalised and determines the impact of the injected kinetic energy fraction. 

We investigated the reliability of numerical solutions of the jet heating problem as a function of the Riemann solver. We found that the HLLE and HLLC solvers produce different results for fixed numerical parameters (resolution, temperature floor) when the system develops small-scale multi-phase structure via thermal instability. Such multi-phase structure also influences the duty cycle of the jet and consequently the heating of the cluster core. It is possible to achieve better agreement between the HLLC and HLLE time averaged solutions by increasing the temperature floor to values $>10\%$ of the virial temperature. This does not however allow full development of a cooling flow or thermal instability. To allow thermal instabilities to develop the temperature floor needs to be kept low ($T_{\rm floor} \leq 5\times 10^5$~K), but this makes the solution more susceptible to the properties of the hydrodynamical solver. Increasing the resolution of future simulations and using solvers with low numerical diffusion may alleviate this problem. Numerical convergence for thermal instability problems is a difficult in part because the full range of scales involved in the process is not understood yet \citep{2018MNRAS.473.5407M}. Furthermore, the chaotic nature of the system may amplify the differences between numerical solutions from different solvers even if thermal instability is appropriately resolved. 

Jet heating in galaxy clusters is a non-linear, chaotic, multi-scale and multi-phase problem which challenges state-of-the-art numerical hydrodynamics codes. Our results complement recent conclusions by \cite{2018arXiv180202177O} who showed that predictions on the mixing of AGN bubbles in simulations can strongly depend on the choice for the hydro solver and its capability to appropriately resolve mixing instabilities. Currently available {hydrodynamical simulations} should be considered as guidelines to interpret the observations, but limitations in numerical models and spatial resolution should be seriously taken into account before drawing quantitative conclusions. Future numerical work on this problem requires (I) low numerical diffusion hydrodynamical methods that capture shocks and contact discontinuities, (II) better spatial resolution to decrease numerical errors, (III) a better theoretical understanding of the scales involved in thermal instabilities in the ICM, (IV) the inclusion of additional physics that may significantly alter the conclusions from pure hydrodynamical simulations \citep[e.g. relativity, magnetic fields, anisotropic thermal conduction and cosmic rays; ][]{2016ApJ...818..181Y, 2017ApJ...844...13R, 2018MNRAS.473.1332G}. 

{Insight on whether currently available hydrodynamical simulations are sufficient to explain the phenomenology of AGN jets in galaxy clusters may come by directly comparing them to observations. X-ray observations of nearby clusters such as Virgo, Perseus or Coma may yield important constraints on the dynamics and thermodynamics of their ICM that has been heated by AGN feedback. Unfortunately, converting X-ray brightness maps and spectra into constraints on the phenomena that heat the cluster core has proven to be difficult. In fact, even for Perseus, one of the most widely studied clusters, constraints on the mechanisms that heat the ICM are based on a series of theoretical assumptions that may result in different interpretations of the observed data \citep[e.g., compare][]{2014Natur.515...85Z, 2016Natur.535..117H, 2017MNRAS.466L..39H, 2017MNRAS.464L...1F}. Combining X-ray, optical and radio data may yield more stringent observational constraints on AGN feedback in clusters for a selected number of object \citep[e.g., ][]{2013ApJ...777..163H}. Furthermore, Faraday rotation measurements can yield additional observational constraints on the density structure of jets/ICM and on the magnetic field strength \citep{2008SSRv..134...93F}, which are needed to understand whether magnetic fields are dynamically relevant in cluster cores. 

The need for additional physics can also be investigated purely on the theoretical side. In fact, the consistency of the subgrid schemes used to launch jets in purely hydrodynamical simulations should be checked against the predictions of relativistic hydrodynamics simulations \citep{2016MNRAS.461L..46T, 2018MNRAS.473.1332G}. The dynamics and thermodynamics of relativistic cosmic rays produced by AGNs in cluster cores is also thought to be relevant, but has only recently been studied with numerical simulations \citep{2013MNRAS.434.2209W, 2017ApJ...844...13R}. Finally, magnetohydrodynamical simulations including anisotropic thermal conduction are needed to establish whether the latter is enhanced or suppressed in a magnetised ICM perturbed by an AGN jet \citep{2011MNRAS.413.1295M}.}

\section*{Acknowledgments}
We thank the reviewer Geoffrey V. Bicknell for his feedback, that allowed us to greatly increase the quality of our paper. We thank Greg Bryan, Yuan Li, Greg Meece, Brian O'Shea, Chris Reynolds, Karen Yang and Mark Voit for their valuable comments on our paper. We also thank Jim Stone for useful conversations. DM was supported in part by the Swiss National Science Foundation postdoctoral fellowship grant P300P2\_161062, in part by NASA ATP grant 12-APT12-0183 and in part by the CTA and DARK-Carlsberg Foundation Fellowship. EQ was supported in part by NASA ATP grant 12-APT12-0183, a Simons Investigator Award from the Simons Foundation and by NSF grant AST-1715070. CAFG was supported by NSF through grants AST-1412836, AST-1517491, AST-1715216, and CAREER award AST-1652522, by NASA through grant NNX15AB22G, by CXO through grant TM7-18007, and by a Cottrell Scholar Award from the Research Corporation for Science Advancement. DF was supported by an NSF Graduate Research Fellowship. The simulations reported in this paper were run and processed on the Savio computer cluster at UC Berkeley and with resources provided by the NASA High-End Computing (HEC) Program through the NASA Advanced Supercomputing (NAS) Division at Ames Research Center (allocations SMD-14-5492, SMD-14-5189, and SMD-15-6530).


\bibliography{main}


\appendix

\section{Total Heating Estimator }\label{appendix:A}
The heating estimator is based on the work of \cite{2015MNRAS.454.1848R}. Using the Lagrangian formalism, the energy equation of ideal fluid dynamics can be expressed 
as a conservation law for the specific entropy $s$:
\begin{equation}\label{eq:energy}
 \rho T \frac{ds}{dt} = H-C
\end{equation}
where $H$ and $C$ are the heating and cooling rates. It is useful to express the specific entropy as a function of the more familiar measure of entropy $K=P/\rho^{\gamma}$:
\begin{equation}\label{eq:entropy_def}
 s = \frac{k_{\rm B}}{\mu m_{\rm p}(\gamma -1)}\log K
\end{equation}
Then entropy conservation becomes:
\begin{equation}\label{eq:entropy-eq}
 \frac{\rho^{\gamma}}{\gamma-1}\frac{dK}{dt}=H-C
\end{equation}
For a purely ideal fluid, in absence of cooling and heating sources ($H=C=0$), entropy conservation is exact:
\begin{equation}
 \frac{dK}{dt}=0.
\end{equation}
Since we are using the conservative code {\sc athena}, it is useful to write the last equation in conservative form:
\begin{equation}\label{eq:entropy_cons}
 \frac{\partial \hat{\kappa}}{dt} + \nabla\cdotp(\hat{\kappa} \vec{v}) = 0 
\end{equation}
where $\hat{\kappa} = \rho K = P/\rho^{(\gamma -1)}$ is a conservative quantity. In our implementation, {\sc athena} solves equation~\ref{eq:entropy_cons} at each time step 
to give the value of $\hat{\kappa}(t)$, the entropy of the fluid in absence of heating and cooling mechanisms at time $t$. The initial condition for equation~\ref{eq:entropy_cons} 
at each time step $t^{(n)}$ is computed by using the conservative solutions of $\rho$ and $P$ at time $t^{(n)}$.  

Our estimate of the heating rate comes from solving for the entropy at a given time $t$ in different ways. 
In our implementation, {\sc athena} simultaneously solves for the fluid equations in conservative form (mass, momentum and energy conservation) and for entropy 
conservation (equation~\ref{eq:entropy_cons}). Due to truncation errors, conservative codes introduce the effect of numerical viscosity in their solutions. 
Even if energy is conserved to machine precision, truncation-level heating is produced by numerical viscosity. This manifests itself as entropy generation: truncation errors 
lead to dissipation of kinetic energy close to the grid scale that is captured as internal energy. This effects mimics what happens in reality, when heating is generated at small 
scales by phenomena like mixing and physical viscosity. The heating estimator is based on this argument.

Let time $t^{(n)}$ to $t^{(n+1)}$ be two subsequent times for which the numerical solution to the fluid equations is computed. At time $t^{(n+1)}$ we'll have density and pressure 
$\rho^{(n+1)}$, $P^{(n+1)}$, respectively. The entropy at the end of the time step will be:
\begin{equation}
 \kappa^{(n+1)} = \frac{P^{(n+1)}}{[\rho^{(n+1)}]^{(\gamma-1)}}
\end{equation}
which is the estimate of entropy predicted by the conservative evolution which includes the contribution from heating processes mediated by numerical viscosity. During the same time step {\sc athena} 
also solves the advection equation for $\hat{\kappa}^{(n+1)}$ (equation~\ref{eq:entropy_cons}), i.e. the entropy expected for the fluid in absence of heating and 
cooling sources. The entropy generated during the time step can be measured as the difference:
\begin{equation}\label{eq:entropy_diff}
 \Delta\kappa^{(n+1)} = \kappa^{(n+1)}-\hat{\kappa}^{(n+1)}.
\end{equation}
If cooling losses are momentarily ignored, this entropy jump can then be converted into a heating rate:
\begin{equation}\label{eq:heating_rate}
 H^{(n+1)} = \frac{[\rho^{(n+1/2)}]^{(\gamma-1)}}{\gamma-1}\frac{\Delta\kappa^{(n+1)}}{\Delta t},
\end{equation}
where $\rho^{(n+1/2)}$ is the density at half time step (readily available in {\sc athena}) and $\Delta t$ is the time step. \cite{2015MNRAS.454.1848R} show that 
the heating rate estimator of equation~\ref{eq:heating_rate} is second order accurate.

In cases in which cooling is also implemented, the treatment described above to estimate heating is still appropriate, because operator splitting can be used. In fact, within a time step several 
operations are done in the following order:
\begin{enumerate}
 \item Hydrodynamics solver + heating estimator.
 \item Subcycling algorithm to include cooling and jet injection.
 \item Iterate from step (i).
\end{enumerate}
In other words, cooling losses/AGN heating are added after the entropy difference (equation~\ref{eq:entropy_diff}) has been computed. At the next time step the hydrodynamics solver will take 
care of the effects of AGN heating and the heating estimator will automatically measure the resulting heating rate in each cell. 

\section{Tests of the Heating Estimator }\label{appendix:B}

The accuracy of the heating estimator has been assessed by performing two tests.

The first test is 1-d strong shock simulation, a variant of a similar test performed by \cite{2015MNRAS.454.1848R}. In this 1-d test, the gas is initialised with homogeneous density $\rho =1$ and pressure $P = 0.01$, 
in the domain $-0.5 \leq x \leq +0.5$. At $x<0$ the gas is initialised with velocity $v_l = 10$ and at $x>0$ the gas is initialised with velocity $v_r = -10$. The system develops two strong shocks that propagate 
leftwards and rightwards at velocities $v_{sh,l} = -0.333$ and $v_{sh,r} = +0.333$. As the shock sweeps more gas, it will heat it up. In regions that have not been reached by the shock this quantity is identically equal to 
zero, because the gas has not been heated up. In regions that have been swept by one shock, the total heating is a non-zero quantity $Q$. At the center of the domain, where the two shocks are generated, the gas 
gets heated twice, by a quantity of heat $2Q$. 

The total heating associated with one shock crossing can be computed by estimating the entropy jump generated by the shock and by using conservation of mass, momentum and energy across the shock. Let the mass, 
momentum, and energy fluxes across the shock be:
\begin{align}
&\dot{m} = \rho v, \\
&\dot{p} = \rho v^2 + P + \tau, \\
&\dot{e} = \frac{1}{2}\rho v^3 + \frac{\gamma}{\gamma -1 } P v + \tau v,
\end{align}
respectively. Here we have added a momentum transfer from viscous dissipation $\tau$ for completeness. It can be verified that the entropy of the gas can be expressed as a function of $\rho$ and of the constants 
$\dot{m}$, $\dot{p}$, and $\dot{e}$:
\begin{equation} \label{eq:K_of_rho}
K(\rho) = \frac{P}{\rho^\gamma} = (\gamma -1 ) \left[ \frac{1}{2} \frac{\dot{m}}{\rho^{\gamma+1}} -\frac{\dot{p}}{\rho^\gamma} + \frac{\dot{e}}{\dot{m}\rho^{\gamma-1}}\right].
\end{equation}
Then total heating can be evaluated by integrating equation~\ref{eq:entropy-eq} in time with $C=0$ (no cooling, for simplicity)
\begin{align} \label{eq:Q_simple}
\nonumber & Q = \int\limits_{pre}^{post} H(t) dt =  \int\limits_{pre}^{post} \frac{\rho^{\gamma}}{\gamma-1}dK = \\
& = \int\limits_{pre}^{post} \frac{\rho^{\gamma}}{\gamma-1} \frac{dK(\rho)}{d\rho} d\rho, 
\end{align}
where in the last equation we have used equation~\ref{eq:K_of_rho}, and where all the integrals are performed with the pre-shock and the post-shock conditions as extremes. If we use the Rankine-Hugoniot 
conditions for a strong shock as the one we are studying, equation~\ref{eq:Q_simple} becomes:
\begin{equation} \label{eq:Q_final}
Q = \rho_{pre} v_{pre}^2 \left[\gamma \log\left( \frac{\gamma+1}{\gamma-1} \right) -2 \right],
\end{equation}
where the subscript $pre$ refers to the gas state before the shock passes in the rest frame of the shock. Equation~\ref{eq:Q_final} can be used to estimate the total heating produced by the shock at each time and 
at each position, provided that the position of the shock is known.  

\begin{figure}
\begin{center}
    \includegraphics[width=0.45\textwidth]{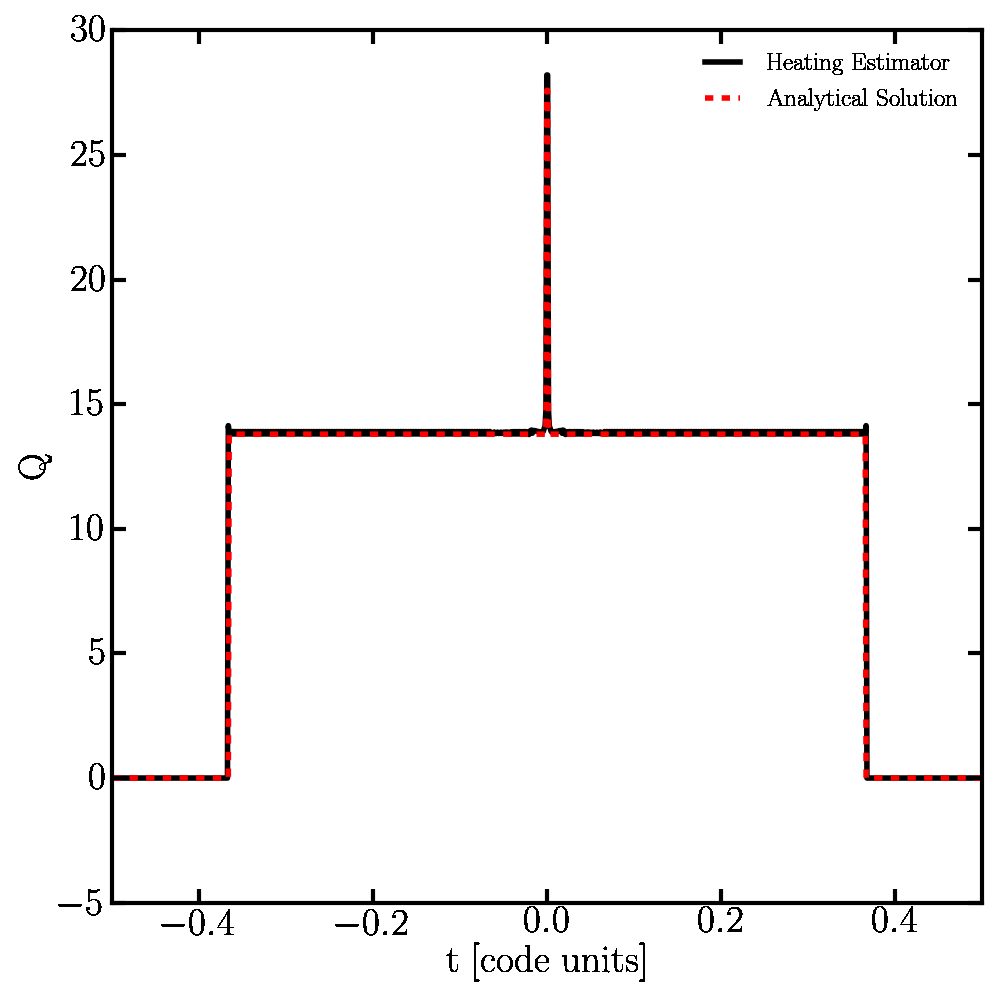}
\end{center}
\caption{ Strong shock test. Time-integrated heating $Q$ as a function of position at time t=0.11 (code units; box size and velocity unit are set to 1). The numerical solution is shown in black, whereas the analytical solution is shown in red. }\label{fig:heat_1dsh}
\end{figure}

In Figure~\ref{fig:heat_1dsh} we show the total heat generated by the shock at all positions at time $t = 0.11$ (code units). The numerical solution provided by our implementation of the heating estimator (black) is 
compared to the analytical solution. The comparison shows excellent agreement for this simple test. 

\begin{figure}
\begin{center}
    \includegraphics[width=0.45\textwidth]{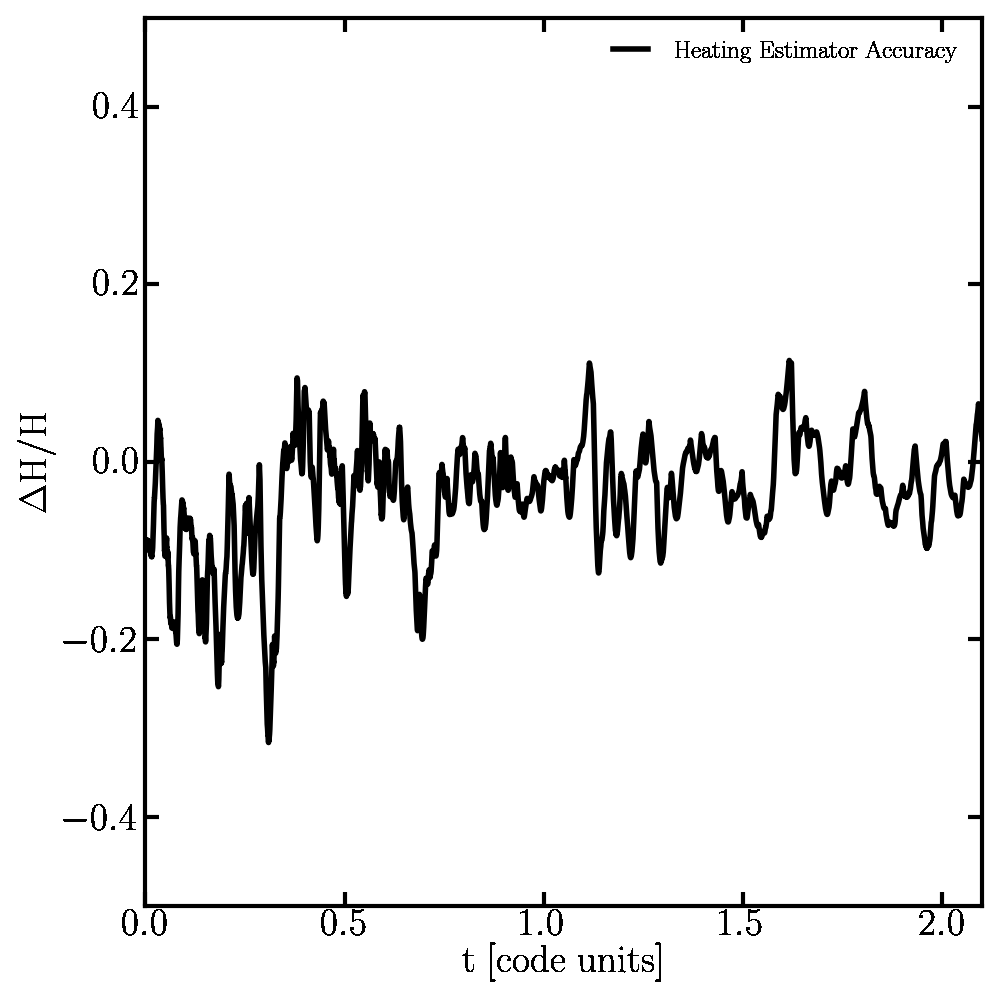}
\end{center}
\caption{ Accuracy of the heating estimator in the periodic turbulent box test. The simulation has been evolved for 21 eddy turnover times. The accuracy of the heating estimator is given by the time average of the line shown in this figure $\langle \Delta H/H \rangle \approx -0.05$.}\label{fig:heat_accuracy}
\end{figure}

The second test is a 3-d box with forced turbulent driving. A cubic box of side 1 with periodic boundary conditions is set up with a gas of uniform density 
($\rho = 10^3$) and temperature ($T=1$). Supersonic velocity perturbations are seeded at each time step to generate turbulent motions with a power spectrum $\propto k^{-2}$, 
with the turbulent driving scale equal to half of the box size. The system is evolved for 21 eddy turnover times. In this case, we know exactly what the heating rate is at each time step and we compare it to the 
estimate provided by our numerical method. For simplicity, we integrate the heating rate over the whole volume. Figure~\ref{fig:heat_accuracy} shows the deviation of the heating estimator $\Delta H/H$ from the instantaneous heating rate
from turbulence, for each time step. The accuracy of the heating estimator is given by the time average of the line shown in Figure~\ref{fig:heat_accuracy}, which is $\langle \Delta H/H \rangle \approx -0.05$
The heating estimator is capable of yielding heating rates with a typical $\sim 5\%$ accuracy in this case. 

\section{Pure Cooling Flow}\label{appendix:C}
{In this Appendix, we show that for a pure cooling flow a direct relation is expected between cooling rate and $PdV$ power. The energy equation for a spherically symmetric cooling flow is
\begin{equation}\label{eq:energy_cf}
\rho T \frac{ds}{dt} = \rho v T \frac{ds}{dr} = -n^2\Lambda = -C,
\end{equation}
which is the same as equation~\ref{eq:energy} without heating sources and with the assumption that the gas only flows radially $v=dr/dt$. Using the definition of the specific entropy $s$ in equation~\ref{eq:entropy_def}, the energy equation becomes:
\begin{equation}\label{eq:energy_cf_trans}
\frac{3}{2}k_{\rm B} n v \frac{dT}{dr}-k_{\rm B}Tv\frac{dn}{dr} = -C.
\end{equation}
Since the gas pressure is $P=nk_{\rm B}T$, if we also assume that the gas participating in the cooling flow is approximately isothermal ($T\approx {\rm const}.$), we conclude that: 
\begin{equation}\label{eq:cf_relation}
v\frac{dP}{dr} \approx C.
\end{equation}
The term of the left hand side of equation~\ref{eq:cf_relation} is the $PdV$ power density. Equation~\ref{eq:cf_relation} demonstrates that for a steady cooling flow, the $PdV$ power is proportional to the cooling rate. 
}

\label{lastpage}
\end{document}